\documentclass[journal]{IEEEtran}

\usepackage{cite}
\usepackage{mathtools}
\usepackage{cuted}
\usepackage{amsmath,amssymb,amsfonts}
\usepackage{pifont}
\usepackage{graphicx}
\usepackage{textcomp}
\usepackage{stfloats}
\usepackage{xcolor}
\usepackage{gensymb}
\usepackage{algorithm}
\usepackage{algorithmic}
\usepackage{geometry}
\usepackage[caption=false,font=footnotesize]{subfig}
\makeatletter
\newcommand\Label[1]{&\refstepcounter{equation}(\theequation)\ltx@label{#1}&}
\makeatother

\def\BibTeX{{\rm B\kern-.05em{\sc i\kern-.025em b}\kern-.08em
    T\kern-.1667em\lower.7ex\hbox{E}\kern-.125emX}}

\newcommand{\tick}{\ding{51}}
\newcommand{\cross}{\ding{55}}
\newtheorem{remark}{Remark}

\belowdisplayskip=\abovedisplayskip
\begin{document}

\title{Second Order Channel Statistics: An Analysis for Optimal Single-user RIS Systems}
\author{Amy~S.~Inwood,~\IEEEmembership{Member,~IEEE,}       
        Peter~J.~Smith,~\IEEEmembership{Fellow,~IEEE,}
        Philippa~A.~Martin,~\IEEEmembership{Senior Member,~IEEE,}
        and~Graeme~K.~Woodward,~\IEEEmembership{Senior Member,~IEEE}
        
    \thanks{This work was presented in part at IEEE GLOBECOM 2024 \cite{inwood_level_2024}.}
    \thanks{A. S. Inwood was with the Department of Electrical and Computer Engineering, University of Canterbury,
    Christchurch, New Zealand, and is now with the Centre for Wireless Innovation (CWI), Queen’s University Belfast, Belfast, BT3 9DT, U.K. (email: a.inwood@qub.ac.uk).}
    \thanks{P. J. Smith is with the School of Mathematics and Statistics, Victoria University of Wellington, Wellington, New Zealand (e-mail: peter.smith@vuw.ac.nz).}
    \thanks{P. A. Martin is with the Department
    of Electrical and Computer Engineering, and G. K. Woodward is with the Wireless Research Centre, University of Canterbury,
    Christchurch, New Zealand (e-mail: (philippa.martin, graeme.woodward)@canterbury.ac.nz).}}

\maketitle

\begin{abstract}
A key challenge facing reconfigurable intelligent surfaces (RISs) is channel state information acquisition. Passive RISs cannot generate pilot signals or process data, making rapid temporal changes in the channel problematic. Additionally, the impact of spatial changes in RIS channels has not been thoroughly investigated. Therefore, in this work, we use second order statistics to investigate the spatio-temporal behaviour of a single-user (SU) RIS system. Assuming a line-of-sight (LoS) RIS to base station (BS) link, we derive an exact expression for the level crossing rate (LCR) of the RIS link (user equipment (UE)-RIS-BS path) and propose a numerically stable approximation for the LCR of the global UE-BS channel. Each LCR expression attained is then utilised to find the corresponding average fade duration (AFD). The temporal signal-to-noise ratio (SNR) correlation is also derived assuming an LoS RIS-BS link. Assuming a Ricean RIS-BS link, expressions for the spatial correlation matrix of the global channel and the mean SNR loss due to channel ageing are derived. All of the analyses are verified by simulation, and the impact of key system parameters is investigated. We show that the use of an RIS does not significantly amplify changes in the channel.
\end{abstract}

\section{Introduction}
Reconfigurable intelligent surfaces (RISs) are an emerging technology for next generation mobile systems that can manipulate the wireless channel with very little power consumption. RISs can enhance several aspects of link performance including signal-to-noise-ratio (SNR) and blockage avoidance \cite{wu_towards_2020}. A considerable body of work on RISs now exists \cite{liu_reconfigurable_2021}, including designs \cite{ibrahim_joint_2023}, measurements \cite{ren_time_2024}, statistical models \cite{dash_ris_2024} and performance analysis \cite{vanchien_outage_2021,cui_snr_2021}. Standards development is also underway, with the European Telecommunications Standards Institute (ETSI) releasing its first report on RISs \cite{etsi_reconfigurable_2023}. Typical RIS implementations involve many elements, $N$, in an array, where each element adjusts the phase of the reflected signal to enhance the channel between the user (UE) and the base-station (BS).

When the RIS changes the channel via its control of the $N$ phase shifts, fundamental properties of the channel may also be changed. These properties include:
\begin{itemize}
	\item Level crossing rate (LCR): The LCR quantifies how often the SNR fading crosses a threshold and is thus used to evaluate temporal changes \cite{ivanis_level_2008, beaulieu_level_2003}.
	\item Average fade duration (AFD): Derived from the LCR, the AFD gives the average length of time the SNR fading spends below a threshold after crossing.
	\item Spatial channel correlation: As the RIS system is comprised of three separate links, the overall spatial channel correlation is a complex function of the parameters in all links. Spatial correlation impacts channel diversity and the ability to perform spatial multiplexing.
	\item SNR correlation: The SNR correlation compares the similarity of the SNR at times $t$ and $t+\tau$, where $\tau$ is a time offset, providing information on how quickly the SNR changes.
\end{itemize}

Passive RIS systems cannot generate pilot signals or process data, making channel estimation difficult \cite{basharat_reconfigurable_2021}. Therefore, rapid changes in the channel are problematic. Temporal second order channel statistics, such as the LCR, AFD, and SNR correlation, can provide insights into how the channel of an RIS system varies with time. While not a second order statistic, channel ageing is another aspect of temporal behaviour that is expected to be important in RIS systems.  If the RIS was designed at time $t$, but the channels have since moved on, the RIS design will become outdated.  The difference between the SNR at times $t$ and $t+\tau$ can indicate how long a fixed RIS design provides performance at an acceptable level.  Additionally, the spatial channel correlation matrix, a key spatial second order statistic, is unknown for an optimal RIS system. These metrics are crucial in providing a comprehensive statistical characterization of the RIS environment. They serve as fundamental analytical tools for validating and understanding channel behavior, quantifying expected performance variations, and understanding how the RIS impacts the underlying wireless propagation environment.

Although some work exists in this area, the complexity of RIS systems has resulted in a limited number of closed-form expressions, restricting analytical insight into system behavior. Work in \cite{girdher_level_2021} and \cite{girdher_second_2023} considers the LCR and AFD of single-user (SU) and multi-user (MU) RIS-only channels, respectively, but neglects any spatial correlation between RIS elements. In \cite{simmons_simulation_2023},  the LCR and AFD of a system with multiple RISs operating cooperatively are investigated, but are necessarily limited to simulation due to the complexity of analysis. In \cite{stefanovic_second_2022}, the LCR and AFD for spatially correlated Fisher-Snedecor channels are derived for multi-hop channels. Work in \cite{anle_double_2023} derives the covariance matrix for a double RIS system, but neglects any direct channel and assumes deterministic RIS phases. Finally, work in \cite{lu_performance_2024, papazafeiropoulos_impact_2023,fang_on_2024} investigates channel ageing in RIS systems, but does not provide closed-form expressions for the loss incurred due to ageing.

In \cite{inwood_level_2024}, for a single input-multiple output (SIMO) single-user (SU) system with an LoS RIS-BS link and correlated Rayleigh direct (user equipment (UE)-base station (BS)) and UE-RIS links, we derived a novel exact analytical expression for the LCR of the RIS-only channel, which occurs when the UE-BS link is blocked. We also proposed a novel, accurate and numerically stable approximation to the LCR of MRC in correlated Rayleigh channels with large numbers of antennas. 

This study presents a significant extension of our earlier work, broadening its scope considerably. We consider correlated Rayleigh direct and UE-RIS channels, and both an LoS and Ricean RIS-BS link. The SU RIS system is considered here as optimal performance can be derived, providing a fundamental benchmark and allowing important mathematical insights to be gained. The phase selection method from \cite{singh_optimal_2021} is used, which is optimal when the RIS-BS link is line-of-sight (LoS). We make the following novel contributions:
\begin{itemize}
    \item For an LoS RIS-BS link:
    \begin{itemize}
        \item We extend the LCR expressions in \cite{inwood_level_2024} to find AFD expressions for the direct and RIS links.
        \item We derive the exact analytical expression for the SNR correlation, which we use to investigate the impacts of Doppler frequency (DF) in an RIS system. 
    \end{itemize}
    \item For a Ricean RIS-BS link:
    \begin{itemize}
        \item We derive the exact analytical expression for the spatial channel correlation matrix. From this, we investigate the sum of channel correlation matrix elements and the percentage of power in leading channel matrix eigenvalues. These provide two useful metrics for the global impact of correlation.
        \item We derive the analytical expression for the mean SNR at time $t+\tau$ relative to the SNR at time $t$ to investigate channel ageing.
        \end{itemize}
\end{itemize}

Analytical results are verified by simulation, and the effects of key system parameters like spatial correlation, DF, system size and channel gain are investigated. These contributions represent a comprehensive and cohesive treatment of the second-order statistics of RIS-assisted systems. To the best of our knowledge, the derived LCR and AFD expressions are the first closed-form results available for spatially correlated RIS channels. For the MRC case, the numerically stable approximations we provide to the LCR and AFD address the known instability of existing exact expressions. We believe this work also presents the first analysis of SNR temporal correlation in a RIS-assisted system, the only derivation of the channel correlation matrix for a single-user RIS with optimal phase design, and the first exact quantification of SNR degradation due to channel aging. Collectively, these results provide a valuable analytical foundation for understanding the dynamic behavior of RIS systems. A comparison of our contributions with existing work in the space of second order statistics of RIS systems is shown in Table \ref{tab:comparison}.

\begin{table*}[t]
  \caption{The contributions of this work compared to existing literature on the second order statistics of RISs.}
    \centering
    \begin{tabular}{|c|c|c|c|c|c|c|c|c|c|c|}
    \hline
        \textbf{Contributions} & \textbf{This Paper} & \cite{inwood_level_2024} & \cite{girdher_level_2021} & \cite{girdher_second_2023} & \cite{simmons_simulation_2023} & \cite{stefanovic_second_2022} & \cite{anle_double_2023} & \cite{lu_performance_2024} & \cite{papazafeiropoulos_impact_2023} & \cite{fang_on_2024}\\
        \hline
         RIS systems & \tick  & \tick & \tick & \tick & \tick & \cross & \tick & \tick & \tick & \tick\\
         \hline
         Spatially correlated channels & \tick & \tick & \cross & \cross & \cross & \tick & \tick & \tick & \tick & \tick \\
         \hline
         Active direct and RIS channels & \tick & \tick & \cross & \cross & \cross  & \cross & \cross & \tick & \tick & \tick\\
         \hline
         Simulation of LCR & \tick & \tick & \tick & \tick & \tick & \tick & \cross & \cross & \cross & \cross \\
         \hline
         Analytical LCR expression & \tick & \tick & \tick & \tick & \cross & \tick & \cross & \cross & \cross & \cross\\
         \hline
         Simulation of AFD & \tick & \tick & \tick & \tick & \tick & \tick & \cross & \cross & \cross & \cross\\
         \hline
         Analytical AFD expression & \tick & \cross &  \tick & \tick & \cross & \tick & \cross & \cross & \cross & \cross\\
         \hline
         Full channel covariance/correlation expression & \tick & \cross & \cross & \cross & \tick & \cross & \tick & \tick & \cross & \cross\\ 
         \hline
         SNR correlation & \tick & \cross & \cross & \cross & \cross & \cross & \cross & \cross & \cross & \cross\\
         \hline
         Simulation of impact of channel ageing & \tick & \cross & \cross & \cross & \cross & \cross & \cross & \tick & \tick & \tick\\
         \hline
         Analytical expression for channel ageing loss &\tick & \cross & \cross & \cross & \cross & \cross & \cross & \cross & \cross & \cross\\
         \hline
    \end{tabular}
    \label{tab:comparison}
\end{table*}

\textit{Notation}: Upper and lower boldface letters represent matrices and vectors, respectively. $\mathbf{v}_k$ is the $k$-th element of $\mathbf{v}$, $\mathbf{v}_{i,k}$ is the $k$-th element of $\mathbf{v}_{i}$ and $\mathbf{M}_{r,s}$ is the $(r,s)$-th element of $\mathbf{M}$. $\mathbb{E}[\cdot]$ represents statistical expectation. $\dot{x}=\frac{d}{dt}x(t)$ refers to the first derivative of $x$ with respect to time. $\mathbb{C}$ is the set of complex numbers. $\mathcal{CN}(\boldsymbol\mu,\mathbf{Q})$ represents a complex Gaussian distribution with mean $\boldsymbol\mu$ and covariance matrix $\mathbf{Q}$. $\chi_k^2$ is a chi-squared distribution with $k$ degrees of freedom. $\mathrm{Exp}(\mu)$ is an exponential distribution with a mean of $\mu$. ${}_1F_1(a,b;z)$ is the confluent hypergeometric function and ${}_2F_1(a,b,c;z)$ is the Gaussian hypergeometric function. $\Gamma(\cdot)$ is the complete gamma function. $K(\cdot)$ and $E(\cdot)$ are the complete elliptic integrals of the first and second kind and $J_0(\cdot)$ is the zeroth order Bessel function of the first kind. $(\cdot)^T$ and $(\cdot)^\dagger$ represent the transpose and Hermitian transpose operators. The angle of a vector, $\mathbf{x}$, of length $N$ is denoted $\angle\mathbf{x}=[\angle{x}_1,\dots,\angle{x}_N]^T$ and the exponent $e^{\mathbf{x}}=[e^{{x}_1},\dots,e^{{x}_N}]^T$. $f_X$ is the probability density function (PDF) of a random variable $X$ and $f_{X,Y}$ is the joint PDF of random variables $X$ and $Y$. $\Phi_X$ is the characteristic function (CF) of  $X$ and $\Phi_{X,Y}$ is the joint CF of  $X$ and $Y$. $\mathrm{Var}(X)$ is the variance of $X$.

\section{System Model}
\begin{figure}[ht]
    \centering
    \includegraphics[scale=0.65]{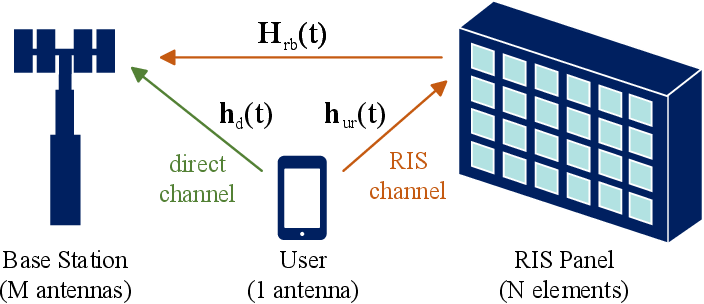}
    \caption{System model showing uplink channels at time $t$.}
    \label{fig:system_diagram}
\end{figure}

This work considers the uplink (UL), RIS-aided system shown in Fig. \ref{fig:system_diagram}. One single antenna user is located near an $N$ element RIS and an $M$ antenna BS.

Let $\mathbf{h}_{\mathrm{d}}(t) \in \mathbb{C}^{M\times 1}$, $\mathbf{h}_{\mathrm{ur}}(t) \in \mathbb{C}^{N\times 1}$ and $\mathbf{H}_{\mathrm{rb}}(t) \in \mathbb{C}^{M\times N}$ be the direct channel, the UE-RIS channel and the RIS-BS channel, respectively, at time instant $t$. We consider the correlated Rayleigh channels $\mathbf{h}_{\mathrm{d}}(t) = \sqrt{\beta_{\mathrm{d}}}\mathbf{R}_{\mathrm{d}}^{1/2}\mathbf{u}_{\mathrm{d}}(t)$ and $\mathbf{h}_{\mathrm{ur}}(t) = \sqrt{\beta_{\mathrm{ur}}}\mathbf{R}_{\mathrm{ur}}^{1/2}\mathbf{u}_{\mathrm{ur}}(t)$. $\beta_{\mathrm{d}}$ and $\beta_{\mathrm{ur}}$ are the channel gains and $\mathbf{R}_{\mathrm{d}}$ and $\mathbf{R}_{\mathrm{ur}}$ are the correlation matrices for the UE-BS link and UE-RIS link, respectively, and $\mathbf{u}_{\mathrm{d}}(t)$ and $\mathbf{u}_{\mathrm{ur}}(t)$ are vectors containing independent and identically distributed (i.i.d) $\mathbb{C}\mathcal{N}(0,1)$ entries. We consider two cases for $\mathbf{H}_{\mathrm{rb}}(t)$, a rank-1 LoS channel, 
\begin{equation}
    \mathbf{H}_{\mathrm{rb}} = \sqrt{\beta_\mathrm{rb}}\mathbf{a}_\mathrm{b}\mathbf
    {a}_\mathrm{r}^\dagger, \label{eq:HrbLoS}
\end{equation}
and a correlated Ricean channel comprised of a rank-1 LoS component and a correlated Rayleigh component,
\begin{equation}
    \mathbf{H}_{\mathrm{rb}}(t) = \sqrt{\beta_\mathrm{rb}}\eta_\mathrm{rb}\mathbf{a}_\mathrm{b}\mathbf
    {a}_\mathrm{r}^\dagger + \sqrt{\beta_\mathrm{rb}}\zeta_\mathrm{rb}\mathbf{R}_{\mathrm{b}}^{1/2}\mathbf{U}_{\mathrm{rb}}(t)\mathbf{R}_{\mathrm{r}}^{1/2},\label{eq:Hrbrice}
\end{equation}
with
\begin{align}
	\eta_\mathrm{rb}=\sqrt{\frac{\kappa_\mathrm{rb}}{\kappa_\mathrm{rb}+1}}, \quad \Label{eq:etarb} &\enskip\! & \zeta_\mathrm{rb}=\sqrt{\frac{1}{\kappa_\mathrm{rb}+1}}. \quad \Label{eq:zetarb} \notag
 \end{align}
 Note that (\ref{eq:Hrbrice}) uses the well-known Kronecker model for correlation. In \eqref{eq:HrbLoS} and \eqref{eq:Hrbrice}, $\beta_\mathrm{rb}$ is the channel gain, $\kappa_\mathrm{rb}$ is the Ricean K-factor, and $\mathbf{a}_\mathrm{b}$ and $\mathbf{a}_\mathrm{r}$ are steering vectors for the LoS ray at the BS and RIS, respectively. $\mathbf{U}^{(k)}_{\mathrm{rb}}(t)$ is a matrix containing i.i.d $\mathbb{C}\mathcal{N}(0,1)$ entries, and $\mathbf{R}_{\mathrm{b}}$ and  $\mathbf{R}_{\mathrm{r}}$ are the correlation matrices for the RIS-BS link at the BS and RIS ends, respectively. The spatio-temporal correlation structure is assumed to be separable as in \cite{smith_impact_2004}, so that
\begin{equation}
    \mathbb{E}\left[\mathbf{h}_{\mathrm{ur},k}(t)\mathbf{h}^*_{\mathrm{ur},l}(t+\tau)\right] = \beta_\mathrm{ur}\mathbf{R}_{\mathrm{ur},k,l}\rho_\mathrm{ur}(\tau),
\end{equation}
\begin{equation}
    \mathbb{E}\left[\mathbf{h}_{\mathrm{d},k}(t)\mathbf{h}^*_{\mathrm{d},l}(t+\tau)\right] = \beta_\mathrm{d}\mathbf{R}_{\mathrm{d},k,l}\rho_\mathrm{d}(\tau).
\end{equation}

$\mathbf{\Phi}(t)\in \mathbb{C}^{N\times N}$ is a diagonal matrix of RIS element reflection coefficients. A design for $\mathbf{\Phi}(t)$ that optimises the total SNR at the BS receiver when $\mathbf{H}_\mathrm{rb}(t)$ is LoS (as in \eqref{eq:HrbLoS}) was proposed in \cite{singh_optimal_2021}, where
\begin{equation}
	\label{eq:Phi}
	\mathbf{\Phi}(t) = \nu(t) \, \mathrm{diag}\left(e^{j\angle \mathbf{a}_{\mathrm{r}}}\right)\mathrm{diag}\left(e^{-j\angle \mathbf{h}_{\mathrm{ur}}(t)}\right),
\end{equation}
and $\nu(t)\!=\!\mathrm{e}^{j\angle{\mathbf{a}_\mathrm{b}^\dagger\mathbf{h}_{\mathrm{d}}(t)}}\!$. Note that when $\mathbf{H}_\mathrm{rb}$ is strongly LoS but has a scattered component, the design is suboptimal, but performs well and is simple to compute. 

The received signal at the BS is
\begin{equation}
    \mathbf{r}(t) = (\mathbf{h}_{\mathrm{d}}(t)+\mathbf{H}_{\mathrm{rb}}(t)\mathbf{\Phi}(t) \mathbf{h}_{\mathrm{ur}}(t))s(t) + \mathbf{n}(t), \label{eq:channel}
\end{equation}
where $s(t)$ is the transmitted signal, $\mathbb{E}[|s_k|^2]=E_s$, and $\mathbf{n}(t)\sim \mathbb{C}\mathcal{N}(0,\sigma^2\mathbf{I})$ is additive white Gaussian noise.

\section{Analysis for a LoS RIS-BS Channel}
In this section we consider a system where the RIS-BS link is LoS as in \eqref{eq:HrbLoS}, and the direct and UE-RIS links are correlated Rayleigh. 
The global UL channel is therefore
\begin{equation}
	\mathbf{h}(t) = \mathbf{h}_\mathrm{d}(t) + \sqrt{\beta_\mathrm{rb}}\nu(t)Y(t)\mathbf{a}_\mathrm{b}, \label{eq:h_LoS}
\end{equation}
where $Y(t) = \sum_{k=1}^{N}|\mathbf{h}_{\mathrm{ur},k}(t)|$. We now investigate the LCR and AFD of the SNR, and the SNR correlation. The more general Ricean case was not considered for all channel statistics as the mathematical complexity would have limited the derivation of closed form expressions, which provide critical insights into system behaviour.

\subsection{Level Crossing Rate}
\label{subsec:LCR}
The level crossing rate of a generic SNR is the rate at which the SNR crosses a threshold, $T$, and is defined in \cite{rice_statistical_1948} as
\begin{equation}
    \mathrm{LCR}(T) = \int\displaylimits_{0}^\infty\dot{x}f_{\mathrm{SNR},\mathrm{\dot{SNR}}}(T,\dot{x})d\dot{x},
\end{equation}
where $f_{\mathrm{SNR},\mathrm{\dot{SNR}}}(T,\dot{x})$ denotes the joint PDF of the SNR and its derivative. Note that the LCR expressions given below are stated for completeness, and their derivations can be found in \cite{inwood_level_2024}. In Sec.~\ref{subsec:LCR} and  Sec.~\ref{subsec:AFD}, results are given for the case where the temporal correlation follows the well-known Jakes' model, i.e., $\rho_\mathrm{d}(\tau)=J_0(2 \pi f_\mathrm{d} \tau)$ and $\rho_\mathrm{ur}(\tau)=J_0(2 \pi f_\mathrm{ur} \tau)$. However, the results only depend on the second derivative of the correlation function. Hence, the results apply to all valid correlation models by replacing the specific values, $f_\mathrm{d}$ and $f_\mathrm{ur}$, by the general expressions, $(-{\rho_\mathrm{d}''(0)}/{2 \pi^2})^{1/2}$ and $(-{\rho_\mathrm{ur}''(0)}/{2 \pi^2})^{1/2}$, respectively.

\subsubsection{LCR of the Direct Channel}
\label{sec:LCRd}
In the absence of an RIS, or when the RIS is blocked, the direct link (setting $\beta_\mathrm{rb}=0$ in (\ref{eq:h_LoS})), has an SNR of
\begin{equation}
    \label{eq:SNRdir}
    \mathrm{SNR}_\mathrm{d}(t) = \frac{E_s}{\sigma^2}\mathbf{h}_{\mathrm{d}}^\dagger(t)\mathbf{h}_{\mathrm{d}}(t).
\end{equation}
The LCR of the SNR in (\ref{eq:SNRdir}) is solved in \cite{ivanis_level_2008} as
\begin{multline}
\label{eq:LCR_direct_exact}
    \mathrm{LCR}_\mathrm{d}(T) = \frac{\sqrt{\pi}(2T)^{\frac{3}{2}}f_\mathrm{d}}{3}\!\sum_{n=1}^M\mathrm{e}^{-\frac{T}{\theta_n}}\sum_{\substack{l=1 \\ l\neq n}}^M\!
    \bigg(\!\frac{\sqrt{\theta_n}}
    {\theta_l(\theta_n\!-\!\theta_l)} \\ \times\!{}_1F_1\!\left(\!1,\tfrac{5}{2};-\tfrac{T}{\theta_l}\right)\!\!\prod\limits_{m \in S_{l,n}}\!\frac{\theta_l\theta_n}{(\theta_l\!-\!\theta_m)(\theta_n\!-\!\theta_m)}\bigg),\!\!\!
\end{multline}
where $S_{l,n} = \{m\in\mathbb{N}\,|\, 1 \leq m \leq M , m\neq n , m\neq l\}$ and $\theta_1...\theta_M$ are the eigenvalues of $\frac{E_s}{\sigma^2}\beta_\mathrm{d}\mathbf{R}_\mathrm{d}$. Note that \cite{ivanis_level_2008} was based on unequal power channels. However, the results can be used here, as $\frac{E_s}{\sigma^2}\mathbf{h}_{\mathrm{d}}^\dagger(t)\mathbf{h}_{\mathrm{d}}(t)=\frac{E_s}{\sigma^2}\beta_{\mathrm{d}} \mathbf{u}_{\mathrm{d}}^\dagger(t)\mathbf{R}_\mathrm{d}\mathbf{u}_{\mathrm{d}}(t)$ is statistically identical to $\mathbf{u}_{\mathrm{d}}^\dagger(t)\mathbf{B}\mathbf{u}_{\mathrm{d}}(t)$, where $\mathbf{B}={\mathrm{diag}}(\theta_1, \ldots ,\theta_M)$, when $\mathbf{u}_{\mathrm{d}}$ is isotropically distributed. This is the case here as $\mathbf{u}_{\mathrm{d}}$ is a vector containing i.i.d. complex Gaussian entries. Hence, the SNR with spatial correlation is statistically identical to the SNR with unequal branch powers. \begin{remark} \label{rem:LCR_precision}
    While (\ref{eq:LCR_direct_exact}) is mathematically correct, it suffers from numerical precision problems. When spatial correlations are small or large, sets of eigenvalues of $\mathbf{R}_\mathrm{d}$ become very similar, and products of tiny eigenvalue differences occur in the denominator of (\ref{eq:LCR_direct_exact}). The process then loses precision. This is more likely to occur when $M \geq 10$.
\end{remark}

Motivated by Remark \ref{rem:LCR_precision}, we proposed a novel numerically stable and accurate approximation for larger MIMO systems in \cite{inwood_level_2024}. This method keeps the first $L$ dominant eigenvalues and replaces the $M-L$ trailing eigenvalues with their average. This allows the LCR to be modelled by a linear combination of exponentials and a single chi-square variable, giving 
\begin{multline}
    \mathrm{LCR_d}(T) \approx \frac{\kappa_0j}{4\pi}\sum_{r=1}^L\bigg(I(r,S,r,1)B_r + \sum_{k=1}^SC_kD_{r,k} \\ \times I(r,S-k+1,L+1,k)\bigg), \label{eq:LCRdfinal}
\end{multline}
where $I(r,S,r,1)$, $B_r$, $C_k$ and $D_{r,k}$ are stated in \cite[(16)-(19)]{inwood_level_2024}.
\begin{remark}
    The LCR expression in \eqref{eq:LCRdfinal} indicates that even in spatially correlated MIMO systems, temporal fading statistics can be analytically captured without full knowledge of every eigenmode, provided dominant modes are preserved.
\end{remark}

\subsubsection{LCR of the RIS Channel}
\label{sec:LCR_RIS}
The SNR for the RIS link (UE-RIS-BS) with no direct path between the transmitter and receiver ($\beta_\mathrm{d}=0$ in (\ref{eq:h_LoS})) is
\begin{equation}
    \mathrm{SNR}_\mathrm{R}(t) = \frac{E_s M\beta_\mathrm{rb}}{\sigma^2}Y^2(t) = cY^2(t), \label{eq:RISonlySNR}
\end{equation}
where $c={E_s M\beta_\mathrm{rb}}/{\sigma^2}$. We derived the LCR of the SNR variable, $\mathrm{SNR}_\mathrm{R}(t) =  cY^2(t)$, across a threshold $T$ in \cite{inwood_level_2024} as
\begin{equation}
     \mathrm{LCR_R}(T) = \sqrt{\tfrac{2}{\pi}\,c\,T\omega^2} f_{\mathrm{SNR_R}}(T), \label{eq:LCR_RIS_exact}
\end{equation}
where $\omega^2 =\pi^2f_\mathrm{ur}^2\beta_\mathrm{ur}\sum_{k=1}^N\sum_{l=1}^N(E(\mathbf{R}_{\mathrm{ur},k,l})-(1\!-\!\mathbf{R}_{\mathrm{ur},k,l}^2)K(\mathbf{R}_{\mathrm{ur},k,l}))$, $K$ and $E$ refer to complete elliptic integrals of the first and second kind and $f_{\mathrm{SNR_R}}(T)$ is the PDF of $\mathrm{SNR_R}$. 
\begin{remark}
    The result in (\ref{eq:LCR_RIS_exact}) has the same general form as the LCR for the SNR of MRC in i.i.d. Rayleigh fading channels \cite{ko_general_2002}, suggesting that RISs do not accentuate temporal variations in the SNR relative to standard MIMO techniques.
\end{remark}

Obtaining a closed-form expression for $\mathrm{LCR_R}(T)$ requires an explicit form for the PDF of $\mathrm{SNR_R}$. We believe an exact expression for $f_\mathrm{SNR_R}(T)$ is intractable. However, $\mathrm{SNR}_\mathrm{R}(t) = cY^2(t)$ and a gamma approximation to $Y(t)$ was given in \cite{singh_optimal_2021} and its accuracy investigated in \cite{kundu}. The gamma approximation is surprisingly accurate over a wide range of typical scenarios. To date, the only cases where the gamma approximation is known to fail is when all the $\mathbf{h}_{\mathrm{ur},k}(t)$ variables are massively correlated or when a few of these variables are dominant. Neither scenario is likely in most RIS scenarios. Using standard transformation theory to compute the approximate probability density function (PDF) of $\mathrm{SNR}_\mathrm{R}(t)$ from the gamma approximation for $Y(t)$ gives the approximate LCR,
\begin{equation}
    \mathrm{LCR_R}(T) \approx \frac{1}{\Gamma(r)}\sqrt{\frac{\omega^2}{2\pi}\,}\,\theta^r\left(\frac{T}{c}\right)^{\frac{r-1}{2}}\!\exp\left(\!-\theta\sqrt{\frac{T}{C}}\right)\!, \label{eq:LCR_RIS}
\end{equation}
where $\theta=\frac{\mathbb{E}[Y(t)]}{\mathrm{Var}[Y(t)]}$ and $r\!=\!\theta\mathbb{E}[Y(t)]$. From \cite{singh_optimal_2021},
\begin{equation}
    \mathbb{E}[Y(t)]=\tfrac{N\sqrt{\pi \beta_\mathrm{ur}}}{2}, \label{eq:EY}
\end{equation}
and
\begin{equation}
    \!\!\!\mathrm{Var[Y(t)]}\!\!=\!\!\frac{\beta_\mathrm{ur}\pi}{4}\!\bigg(\!\sum^N_{i=1}\!\sum_{j\neq i}\!{}_2F_1\!\!\left(\!-\tfrac{1}{2}\!,\!-\tfrac{1}{2}\!,\!1,\!|\mathbf{R}_{\mathrm{ur},i,j}|^2\right)\!\!-\!N^2\!\!\bigg)\!.\!
    \label{eq:varY}
\end{equation}

\subsection{Average Fade Duration}
\label{subsec:AFD}
The AFD is the average amount of time the SNR spends below a threshold, $T$, in one fade, and is defined in terms of the LCR and $F_\mathrm{SNR}(T)$ (the cumulative distribution function (CDF) of SNR) as \cite{rice_statistical_1948}
\begin{equation}
    \mathrm{AFD}(T) = \frac{F_\mathrm{SNR}(T)}{\mathrm{LCR}(T)}. \label{eq:AFD}
\end{equation}

\subsubsection{AFD of the Direct Channel}
The direct link SNR in (\ref{eq:SNRdir}) can be represented by the PDF of a hypoexponential distribution,
\begin{equation}
    f_\mathrm{SNR_d}(T) = \sum_{i=1}^M \prod_{j\neq i}\left(\frac{\theta_i}{\theta_i-\theta_j}\right)\frac{1}{\theta_i}\mathrm{e}^{-\frac{T}{\theta_i}}.
\end{equation}
Integrating this expression gives the CDF,
\begin{equation}
    F_\mathrm{SNR_d}(T) \!=\!\int_0^T \!\!\! f_\mathrm{SNR_d}(u) du\!=\!1\!-\! \sum_{i=1}^M \prod_{j\neq i}\left(\frac{\theta_i}{\theta_i-\theta_j}\right)\mathrm{e}^{\frac{-T}{\theta_i}}. \notag
\end{equation}
Like the exact LCR in (\ref{eq:LCR_direct_exact}), the difference of eigenvalues in the denominator causes precision problems. The approximation to the SNR CDF proposed in \cite{smith_on_2014} is applied, so that
\begin{equation}
    F_\mathrm{SNR_d}(T)\approx \frac{1}{2} - \sum_{t=0}^L\frac{\Im(\Psi((t+\frac{1}{2})\Delta)\mathrm{e}^{-j(t+\frac{1}{2})\Delta T})}{\pi(t+\frac{1}{2})}, \notag
\end{equation}
where $\Psi(s) = \mathbb{E}[\mathrm{e}^{jsX}]$ is the characteristic function of $X = \sum_{i=1}^M\theta_iY_i$ and $Y_i\sim\mathrm{Exp}(1)$. Thus
\begin{align}
    \Psi(s) &= \mathbb{E}\bigg[\mathrm{exp}\bigg(\sum_{i=1}^M js\theta_iY_i\bigg)\bigg]=\prod_{i=1}^M\left(\frac{1}{1-js\theta_i}\right).
\end{align}
$\Delta$ is chosen so $P(X<T-\frac{2\pi}{\Delta})\approx 0$ and $P(X>T+\frac{2\pi}{\Delta})\approx 0$, and $L$ is increased until the result converges. Therefore,
\begin{equation}
    F_\mathrm{SNR_d}(T)\approx \frac{1}{2} - \sum_{t=0}^L\frac{\Im\!\left(\mathrm{e}^{-j(t+\frac{1}{2})\Delta T}\prod_{i=1}^M\left(\frac{1}{1-j\theta_i(t+\frac{1}{2})\Delta}\right)\!\right)}{\pi(t+\frac{1}{2})}. \notag
\end{equation}

Combining this and (\ref{eq:LCRdfinal}) into (\ref{eq:AFD}) gives an accurate approximation to the AFD for the direct channel. Note that the approximation solely resides in the truncation of $F_\mathrm{SNR_d}(T)$ to a finite sum. All other calculations are exact as shown in \cite{davies}. 
\begin{remark}
     Increasing spatial correlation leads to higher AFD values at low SNR thresholds. Eigenvalue clustering reduces the variability across diversity branches, increasing the probability of sustained low-SNR conditions while simultaneously reducing the rate of threshold crossings. 
\end{remark}

\subsubsection{AFD of the RIS Channel}
As stated in Sec. \ref{sec:LCR_RIS}, the SNR PDF (and therefore the CDF) are intractable, so again, the accurate gamma approximation will be used instead. Therefore, we can write the approximate PDF as
\begin{equation}
	f_\mathrm{SNR_R}(T) \approx \frac{\theta^r}{2\Gamma(r)}\frac{T^{r/2-1}}{c^{r/2}}\mathrm{e}^{-\theta \sqrt{T/c}}, \label{eq:gammaPDFSNR}
\end{equation}
where $\theta$, $r$, $\mathbb{E}[Y(t)]$ and $\mathrm{Var}[Y(t)]$ are as given in Sec. \ref{sec:LCR_RIS}. Thus, integrating (\ref{eq:gammaPDFSNR}) approximates the CDF as
\begin{equation}
    F_\mathrm{SNR_R}(T) \approx \frac{1}{\Gamma(r)}\,\gamma\left(r,\theta \sqrt{\tfrac{T}{c}}\right). \label{eq:gammaCDFSNR}
\end{equation}
Combining (\ref{eq:gammaCDFSNR}), (\ref{eq:LCR_RIS}) and (\ref{eq:AFD}) provides a reliable estimate to the AFD of the RIS channel. 
\begin{remark}
    The RIS AFD expression shares the same mathematical structure as that of conventional MIMO channels, with the threshold scaling governed by the average and variance of summed amplitudes. This confirms that RISs do not induce additional temporal instability in fade behavior.
\end{remark}

\subsection{SNR Correlation}
In this section, we derive an exact expression for the temporal SNR correlation at times $t$ and $t+\tau$ of a full RIS system (all links operating). Note that the results here and in Sec. \ref{sec:analysisRicean} are given in terms of generic correlation functions, $\rho_\mathrm{d}(\tau)$ and $\rho_\mathrm{ur}(\tau)$.  Hence, they are fully general and apply for any valid temporal correlation function. The correlation of the SNR at times $t$ and $t+\tau$ is

\begin{equation}
	\rho_\mathrm{SNR} = \frac{\mathbb{E}\left[\mathrm{SNR}(t) \mathrm{SNR}(t+\tau)\right]-\mathbb{E}\left[\mathrm{SNR}(t)\right]^2}{\mathbb{E}\left[(\mathrm{SNR}(t))^2\right]-\mathbb{E}\left[\mathrm{SNR}(t)\right]^2}. \label{eq:SNRcorr_SOS}
\end{equation}
Let $\epsilon = \mathbb{E}\left[\mathrm{SNR}(t) \mathrm{SNR}(t+\tau)\right]$. Using (\ref{eq:h_LoS}),
\begin{equation}
    \mathrm{SNR}(t) = a(t) + 2b(t) + c(t),
\end{equation}
where $a(t) = \mathbf{h}_\mathrm{d}^\dagger(t)\mathbf{h}_\mathrm{d}(t)$, $b(t) = \sqrt{\beta_{\mathrm{rb}}}Y(t)|\mathbf{a}_\mathrm{b}^\dagger\mathbf{h}_{\mathrm{d}}(t)|$, and $c(t) = M\beta_\mathrm{rb}Y(t)^2$. As $\mathbb{E}[a(t)b(t+\tau)] = \mathbb{E}[b(t)a(t+\tau)]$, $\mathbb{E}[a(t)c(t+\tau)] = \mathbb{E}[c(t)a(t+\tau)]$ and $\mathbb{E}[b(t)c(t+\tau)] = \mathbb{E}[c(t)b(t+\tau)]$, $\epsilon$ is given by
\begin{multline}
    \epsilon = \mathbb{E}\big[a(t)a(t+\tau) + 4a(t)b(t+\tau) + 2a(t)c(t+\tau) \\
    + 4b(t)b(t+\tau) + 4b(t)c(t+\tau) + c(t)c(t+\tau)\big].\!
     \label{eq:SNR_correlation}
\end{multline}
Note that the equality of cross-time expectations, e.g., $\mathbb{E}[a(t) b(t+\tau)] = \mathbb{E}[b(t) a(t+\tau)]$, relies on the assumption that the involved stochastic processes are wide-sense stationary. This ensures that the joint second-order statistics  are symmetric and depend solely on the time difference, $\tau$. Derived in Appendix A, the six terms in (\ref{eq:SNR_correlation}) are
\begin{multline}
    \mathbb{E}[a(t)a(t+\tau)] = \beta_\mathrm{d}^2(|\rho_\mathrm{d}(\tau)|^2(M^2 + \mathrm{tr}(\mathbf{R}_\mathrm{d}^2)) \\ + M^2(1-|\rho_\mathrm{d}(\tau)|^2)), \label{eq:Eatatau}
\end{multline}
\begin{multline}
    \mathbb{E}[a(t)b(t+\tau)] = \frac{N\pi\beta_\mathrm{d}\sqrt{\beta_\mathrm{d}\beta_\mathrm{rb}\beta_\mathrm{ur}\mathbf{a}_\mathrm{b}^\dagger\mathbf{R}_\mathrm{d}\mathbf{a}_\mathrm{b}}}{4} \\
    \times\bigg(M+|\rho_\mathrm{d}(\tau)|^2\frac{\mathbf{a}_\mathrm{b}^\dagger\mathbf{R}_\mathrm{d}^2\mathbf{a}_\mathrm{b}}{2\mathbf{a}_\mathrm{b}^\dagger\mathbf{R}_\mathrm{d}\mathbf{a}_\mathrm{b}}\bigg), \label{eq:Eatbtau}
\end{multline}
\begin{multline}
    \mathbb{E}[a(t)c(t+\tau)] = M^2\beta_\mathrm{d}\beta_\mathrm{rb}\beta_\mathrm{ur}\bigg(N + \frac{\pi}{4}\sum_{i=1}^N\sum_{j\neq i} \\ \times{}_2F_1\!\left(\!-\tfrac{1}{2},-\tfrac{1}{2},1,|\mathbf{R}_{\mathrm{ur},i,j}|^2\right)\!\!\!\bigg), \label{eq:Eatctau}
\end{multline}
\begin{multline}
    \!\!\mathbb{E}[b(t)b(t+\tau)] {=} \frac{\pi^2\!\beta_\mathrm{d}\beta_\mathrm{rb}\beta_\mathrm{ur}\mathbf{a}_\mathrm{b}\!\mathbf{R}_\mathrm{d}\mathbf{a}_\mathrm{b}}{16}{}_2F_{1}\!\left(\!-\tfrac{1}{2}\!,\!-\tfrac{1}{2}\!,\!1,\!|\rho_\mathrm{d}\!(\tau)|^2\right)\\\times\sum_{i=1}^N\sum_{j\neq i}\!{}_2F_1\!\left(-\tfrac{1}{2},-\tfrac{1}{2},1,|\rho_\mathrm{ur}(\tau)|^2|\mathbf{R}_{\mathrm{ur},i,j}|^2\right)\!,\!\! \label{eq:Ebtbtau}
\end{multline}
\begin{multline}
    \mathbb{E}[b(t)c(t+\tau)]= \frac{M\beta_\mathrm{rb}\sqrt{\pi\beta_\mathrm{d}\beta_\mathrm{rb}\,\mathbf{a}_\mathrm{b}^\dagger\mathbf{R}_\mathrm{d}\mathbf{a}_\mathrm{b}}}{2} \\ \times \alpha^3r(r+2)(4\gamma^2+r), \label{eq:Ebtctau}
\end{multline}
\begin{multline}
        \mathbb{E}[c(t)c(t+\tau)] = M^2\beta_\mathrm{rb}^2\alpha^4 r (r+2) \\ \times(r^2 + 8r\gamma^2 + 2r + 8\gamma^4 + 16\gamma^2), \label{eq:Ectctau}
\end{multline}
where 
\begin{equation}
    \gamma = \frac{\sum\limits_{i=1}^N\sum\limits_{j=1}^N{}_2F_1\!\!\left(-\frac{1}{2},-\frac{1}{2},1,|\rho_\mathrm{ur}(\tau)|^2|\mathbf{R}_{\mathrm{ur},i,j}|^2\right)\!-\!N^2}{\sum\limits_{i=1}^N\sum\limits_{j=1}^N{}_2F_1\left(-\frac{1}{2},-\frac{1}{2},1,|\mathbf{R}_{\mathrm{ur},i,j}|^2\right)\!-\!N^2}\!,\!\!\! \label{eq:rho}
\end{equation}
and $\alpha = \frac{\mathrm{Var}[Y(t)]}{2\mathbb{E}[Y(t)]}$, $r=\frac{\mathbb{E}[Y(t)]}{\alpha}$, and $\mathbb{E}[Y(t)]$ and $\mathrm{Var}[Y(t)]$ are given in (\ref{eq:EY}) and (\ref{eq:varY}), respectively. $\mathbb{E}[\mathrm{SNR}(t)]$ is given in (6) of \cite{singh_optimal_2021}, and $\mathbb{E}[\left(\mathrm{SNR}(t)\right)^2]$ can be found by setting $\tau = 0$ in (\ref{eq:SNR_correlation}). Therefore, (\ref{eq:SNRcorr_SOS}) is fully known.
\begin{remark}
\label{rem:SNRcorr}
    Due to the quadratic nature of the SNR (i.e. $\mathrm{SNR}(t)\propto||\mathbf{h}(t)||^2$), correlation remains strictly positive, even when the underlying channel exhibits negative correlation. This indicates that any temporal correlation is helpful, including negatively correlated channels.
\end{remark}

\section{Analysis for a Correlated Ricean RIS-BS Channel} \label{sec:analysisRicean}
In this section we consider a system where the RIS-BS link is Ricean as in \eqref{eq:Hrbrice}, and the direct and UE-RIS links are correlated Rayleigh. Letting $c_1 = \sqrt{\beta_\mathrm{rb}}\eta_\mathrm{rb}$, $c_2 = \sqrt{\beta_\mathrm{rb}}\zeta_\mathrm{rb}$ and $\mathbf{G}(t)=\mathbf{R}_\mathrm{b}^{1/2}\mathbf{U}_\mathrm{rb}(t)\mathbf{R}_\mathrm{r}^{1/2}$, the global UL channel is  
\begin{equation}
    \!\!\!\mathbf{h}(t)\!\!=\!\!\mathbf{h}_\mathrm{d}(t)\!{+}c_1\nu(t)Y\!(t)\mathbf{a}_\mathrm{b}\!{+}c_2\mathbf{G}(t)\nu(t)\mathrm{diag}(\mathbf{a}_\mathrm{r})|\mathbf{h}_\mathrm{ur}(t)|.\!\! \label{eq:chanrice}
\end{equation}
We now investigate the channel correlation matrix and the SNR loss due to channel ageing.

\subsection{Channel Correlation Matrix}
\label{sec:chancorr} As $\mathbf{h}_\mathrm{d}(t)$ is a correlated Rayleigh channel, $\mathbb{E}[\mathbf{h}_\mathrm{d}(t)] \!=\!\mathbb{E}[\nu(t)]\!=\! 0$, and thus $\mathbb{E}[\mathbf{h}(t)] = 0$. Therefore, the channel correlation matrix, $\mathbf{R}_\mathrm{h}$, at time $t$ is
\begin{equation}
	\mathbf{R}_\mathrm{h}(t) = \mathrm{diag}\left(\textrm{\boldmath${\psi}$}(t)\right)\mathbf{D}(t)\,\mathrm{diag}\left(\textrm{\boldmath${\psi}$}(t)\right), \label{eq:channcorr}
\end{equation}
where $\mathbf{D}(t) = \mathbb{E}\left[\mathbf{h}(t)\mathbf{h}(t)^\dagger\right]$. The $i$-th element of vector \boldmath${\psi}$\unboldmath$(t)$ is \boldmath${\psi}$\unboldmath${}_i(t) = 1/{\sqrt{\mathbb{E}\left[|\mathbf{h}_i(t)|^2\right]}}$, where $\mathbf{h}_i$ is the $i$-th element of $\mathbf{h}$. All terms are at the same time instant so time notation is dropped. From the results in (29) - (32) of \cite{inwood_phase_2023},
\begin{align}
	\mathbb{E}&\left[|\mathbf{h}_i|^2\right]=\beta_\mathrm{d} + \frac{N\pi\eta_\mathrm{rb}\sqrt{\beta_{\mathrm{d}}\beta_{\mathrm{rb}}\beta_{\mathrm{ur}}}}{2}\,\Re\left\{\frac{\mathbf{a}^*_{\mathrm{b},i}\mathbf{R}_{\mathrm{d},i}\mathbf{a}_\mathrm{b}}{\sqrt{\mathbf{a}_\mathrm{b}\mathbf{R}_\mathrm{d}\mathbf{a}_\mathrm{b}}}\right\}\notag \\ & + \beta_\mathrm{rb}\beta_\mathrm{ur} \bigg[N\left(\eta_\mathrm{rb}^2+\zeta_\mathrm{rb}^2\right) + \frac{\pi}{4}\sum_{l=1}^{N}\sum_{j\neq l}\big(\eta_\mathrm{rb}^2 +\zeta_\mathrm{rb}^2\notag \\ &\times\mathbf{a}^*_{\mathrm{r},l}\mathbf{R}_{\mathrm{r},l,j}\mathbf{a}_{\mathrm{r},j}\big){}_2F_{1}\!\!\left(-\tfrac{1}{2},-\tfrac{1}{2},1;\left|\mathbf{R}_{\mathrm{ur},l,j}\right|^2\right)\!\!\bigg], \label{eq:Ehi2}
\end{align}
where $\mathbf{R}_{\mathrm{d},i}$ denotes the $i$-th column of $\mathbf{R}_\mathrm{d}$.
\begin{remark}
    Although most terms in (\ref{eq:Ehi2}) are independent of $i$, the term $\Re\{\mathbf{a}_{\mathrm{b},i}^*\mathbf{R}_{\mathrm{d},i}\mathbf{a}_\mathrm{b}\}$ varies. Hence, the optimal RIS channel has the unusual property that the $M$ end-to-end channels have different powers, due to the varying alignment between $\mathbf{a}_\mathrm{b}$ and the $i$-th row of $\mathbf{R}_\mathrm{d}$. The RIS seeks to exploit these alignments to improve the SNR, but not all are equal.
\end{remark}

We now proceed with the derivation of the channel covariance matrix, $\mathbf{D}$. As $\mathbf{h}_\mathrm{d}$ and $\mathbf{h}_\mathrm{ur}$ are independent correlated Rayleigh channels, and $\mathbb{E}\left[\mathbf{h}_\mathrm{d}\right] = \mathbb{E}\left[\mathbf{h}_\mathrm{ur}\right] = 0$, $\mathbf{D}$ expands to
\begin{multline}
    \mathbf{D} = \mathbb{E}[\mathbf{h}_\mathrm{d}\mathbf{h}_\mathrm{d}^\dagger] + 2\Re(c_1\mathbb{E}[\nu^*\mathbf{h}_\mathrm{d}]\mathbb{E}[Y]\mathbf{a_\mathrm{b}}^\dagger) + c_1^2\mathbb{E}[Y^2]\mathbf{a}_\mathrm{b}\mathbf{a}_\mathrm{b}^\dagger \\ + c_2^2\mathbb{E}[\,\mathbf{G}\,\mathrm{diag}(\mathbf{a}_\mathrm{r})|\mathbf{h}_\mathrm{ur}||\mathbf{h}_\mathrm{ur}|^T\mathrm{diag}(\mathbf{a}_\mathrm{r})\mathbf{G}^\dagger]. \label{eq:Dsimplified}
\end{multline}
As $\mathbf{h}_\mathrm{d}$ has the correlation matrix $\mathbf{R}_\mathrm{d}$,
\begin{equation}
    \mathbb{E}[\mathbf{h}_\mathrm{d}\mathbf{h}_\mathrm{d}^\dagger] = \beta_\mathrm{d}\mathbf{R}_\mathrm{d}. \label{eq:hdhddag}
\end{equation}
From \cite{singh_optimal_2021}, $\mathbb{E}[Y]$ is as stated in (\ref{eq:EY}) and 
\begin{equation}
    \mathbb{E}[Y^2] = \beta_\mathrm{ur}\bigg(\!N + \frac{\pi}{4}\sum_{i=1}^N\sum_{j\neq i}{}_2F_1\!\left(-\tfrac{1}{2},-\tfrac{1}{2},1,|\mathbf{R}_{\mathrm{ur},i,j}|^2\right)\!\!\!\bigg)\!. \notag
\end{equation}
The remaining two terms are derived in Appendix B to be
\begin{equation}
    \mathbb{E}[\nu^*\mathbf{h}_\mathrm{d}] = \frac{\sqrt{\beta_\mathrm{d}\pi}\mathbf{R}_\mathrm{d}\mathbf{a}_\mathrm{b}}{2\sqrt{\mathbf{a}_\mathrm{b}\mathbf{R}_\mathrm{d}\mathbf{a}_\mathrm{b}}},
\end{equation}
and
\begin{align}
    \mathbf{Z} &= \mathbb{E}[\,\mathbf{G}\,\mathrm{diag}(\mathbf{a}_\mathrm{r})|\mathbf{h}_\mathrm{ur}||\mathbf{h}_\mathrm{ur}|^T\mathrm{diag}(\mathbf{a}_\mathrm{r})\mathbf{G}^\dagger], \notag \\ &= \mathbf{R}_\mathrm{b}\mathrm{tr}(\mathrm{diag}(\mathbf{a}_\mathrm{r})\mathbf{B}\mathrm{diag}(\mathbf{a}_\mathrm{r}^*)\mathbf{R}_\mathrm{r}),
\end{align}
where the $i,j$-th element of $\mathbf{B}$ can be written as 
\begin{equation}
	\mathbf{B}_{i,j} = \beta_\mathrm{ur}\frac{\pi}{4}{}_2F_1\left(-\tfrac{1}{2},-\tfrac{1}{2},1;|\mathbf{R}_{\mathrm{ur},i,j}|^2\right).\label{eq:Bij}
\end{equation}
Therefore, combining (\ref{eq:Ehi2}) and (\ref{eq:Dsimplified}) - (\ref{eq:Bij}) with (\ref{eq:channcorr}) gives the channel correlation matrix.

\begin{remark}
\label{rem:uncorr}
When the scattered channel components are uncorrelated, the presence of a line-of-sight (LoS) component in the RIS-BS link (\(\kappa_{\mathrm{rb}} \ne 0\)) prevents the channel covariance matrix from becoming fully diagonal. This is due to the rank-1 structure of the LoS contribution,
$\Sigma^{(\mathrm{LoS})} = c_1 \mathbb{E}[Y^2]\, \mathbf{a}_{\mathrm{b}} \mathbf{a}_{\mathrm{b}}^\dagger$,
which introduces persistent spatial correlation across the receive antennas.
\end{remark}
\begin{remark}
\label{rem:corr}
    Also when $\kappa_\mathrm{rb} \neq 0$, unless the RIS is perfectly broadside to the BS, the channel is never fully correlated. Even if the scattered components of the channel become perfectly correlated, the LoS components contain different phase shifts, i.e. $\mathbf{a}_{\mathrm{b},i} \neq \mathbf{a}_{\mathrm{b},j}$, and thus are not identical, preventing $\mathbf{R_\mathrm{h}}(t)$ from becoming a matrix of ones.
\end{remark}
\subsection{Ageing Loss}
\label{sec:ageing_loss}
Assuming perfect tracking of an end-to-end channel with an outdated RIS design, the channel at time $t+\tau$ is
\begin{equation}
    \mathbf{h}(t+\tau)\! =\! \mathbf{h}_\mathrm{d}(t+\tau) +(c_1\mathbf{a}_\mathrm{b}\mathbf{a}_\mathrm{r}^\dagger + c_2\mathbf{G}(t+\tau))\mathbf{\Phi}(t)\mathbf{h}_\mathrm{ur}(t+\tau). 
    \label{eq:httau}
\end{equation}
The channel ageing loss can be calculated by
\begin{equation}
    L = \mathbb{E}[\mathrm{SNR}(t)-\mathrm{SNR}(t+\tau)].
\end{equation}
First derived in \cite{inwood_phase_2023}, $\mathbb{E}[\mathrm{SNR}(t)]$ is given by
\begin{align}
    &\mathbb{E}[\mathrm{SNR}(t)] {=}\frac{E_s}{\sigma^2}\!\bigg[\!M\!\beta_\mathrm{d}{+}\frac{N\pi\eta_\mathrm{rb}\sqrt{\!\beta_\mathrm{d}\beta_\mathrm{rb}\beta_\mathrm{ur}}||\mathbf{R}_\mathrm{d}^{\!1/2}\!\mathbf{a}_\mathrm{b}||}{2}{+} M\!\frac{\pi}{4} \!\beta_\mathrm{rb}\notag \\ &\!\!\times\!\beta_\mathrm{ur}\!\left(\!\eta_\mathrm{rb}^2\!{+}\zeta_\mathrm{rb}^2\mathrm{tr}\!\left(\!\mathrm{diag}(\mathbf{a}_\mathrm{r}^\dagger)\mathbf{R}_\mathrm{r}\mathrm{diag}(\mathbf{a}_\mathrm{r}){}_2F_1\!\left(\!\!-\tfrac{1}{2}\!,\!-\tfrac{1}{2}\!,\!1,\!|\mathbf{R}_\mathrm{ur}|^2\right)\!\right)\!\right)\!\!\bigg]\!. \notag
\end{align}
$\mathbb{E}[\mathrm{SNR}(t+\tau)]$ is derived in Appendix C and stated in \eqref{eq:ESNRtplustau} at the top of the following page, and $\mathbf{M}$ is a $N\times N$ matrix with $i,j$-th element $\mathbf{M}_{i,j} = {}_2F_1\left(-\frac{1}{2},-\frac{1}{2},1;|\mathbf{R}_{\mathrm{ur},i,j}|^2\right)$.
\begin{figure*}[t]
\normalsize
\begin{align}
     \mathbb{E}[\mathrm{SNR}(t\!+\!\tau)]\!&=\! \frac{E_s}{\sigma^2}\!\bigg[\!M\beta_\mathrm{d}\!+\! \frac{N\pi\eta_\mathrm{rb}\sqrt{\beta_\mathrm{d}\beta_\mathrm{rb}\beta_\mathrm{ur}\mathbf{a}_\mathrm{b}^\dagger\mathbf{R}_\mathrm{d}\mathbf{a}_\mathrm{b}}}{4}(\rho_\mathrm{d}^*(\tau)\rho_\mathrm{ur}(\tau)\!+\!\rho_\mathrm{d}(\tau)\rho_\mathrm{ur}^*(\tau)\!)\!+\! M\beta_\mathrm{rb}\beta_\mathrm{ur}\eta^2_\mathrm{rb}|\rho_\mathrm{ur}(\tau)|^2\bigg(\!N\!+\!\frac{\pi}{4}\notag \\ &\times\!\sum_{i=1}^N\sum_{j\neq i}\!{}_2F_1\!\left(\!-\tfrac{1}{2},-\tfrac{1}{2},1,|\mathbf{R}_{\mathrm{ur},i,j}|^2\right)\!\!\!\bigg)\!\! +\! \!M\beta_\mathrm{rb}\eta_\mathrm{rb}^2 \frac{\pi}{4}\!\big(1\!-\!|\rho_\mathrm{ur}(\tau)|^2\big)\!\!\sum^N_{i=1}\sum^N_{j=1}(\mathbf{R}_{\mathrm{ur},i,j})^2{}_2F_1\!\left(\tfrac{1}{2},\tfrac{1}{2},2,|\mathbf{R}_{\mathrm{ur},i,j}|^2\right) \notag \\ & + M\beta_\mathrm{rb}\beta_\mathrm{ur}\zeta_\mathrm{rb}^2\bigg(\!|\rho_\mathrm{ur}(\tau)|^2\,\mathrm{tr}\bigg(\frac{\pi}{4}\,\mathrm{diag}(\mathbf{a}_\mathrm{r}^*)\,\mathbf{R}_\mathrm{r}\,\mathrm{diag}(\mathbf{a}_\mathrm{r})\mathbf{M}\!\!\bigg)+\!\big(1\!-\!|\rho_\mathrm{ur}(\tau)|^2\big)\mathrm{tr}\big(\mathbf{R}_\mathrm{ur}\mathbf{D}\big)\!\!\bigg)\!\bigg]. \label{eq:ESNRtplustau}
\end{align}
\hrulefill
\end{figure*}
\begin{remark}
    The degradation due to channel ageing is more pronounced when the RIS-BS link exhibits a strong line-of-sight (LoS) component (\(\kappa_{\mathrm{rb}}\) is large) and the number of RIS elements \(N\) is high. In such cases, the RIS contributes more significantly to the overall SNR, making the system more sensitive to misalignment caused by outdated phase shifts.
\end{remark}

\section{Numerical Results}
\label{sec:num_res}
\begin{figure}[ht]
    \centering
    \includegraphics[trim={0cm 0cm 0cm 0cm},clip,scale=0.65]{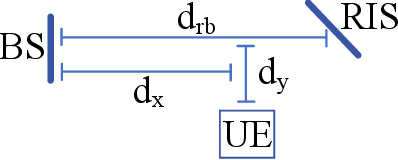}
    \caption{Plan view of the simulation system layout \cite{wu_intelligent_2019}.}
    \label{fig:power_layout}
\end{figure}

Numerical results verify the analyses above and explore the impact of a range of system parameters. For all simulations, $10^6$ replicates were generated. The channel gain values were selected using the distance based path loss model in \cite{wu_intelligent_2019}, where $\beta =  C_0\left({d}/{D_0}\right)^{-\alpha}$, $D_0$ is the reference distance of 1 m, $C_0$ is the path loss at $D_0$ (-30 dB), $d$ is the link distance in metres and $\alpha$ is the path loss exponent. Setting $d_\mathrm{rb}$, $d_x$ and $d_y$ in Fig. \ref{fig:power_layout} gives the planar link distances. The RIS and the BS are set at a height of 15~m. All UEs are assumed to be at a height of 1.5~m, however, this does not impact the Rayleigh fading model used for all UE links. The four layouts used for simulation are detailed in Table \ref{tab:syslayouts}.

\begin{table}[ht!]
\centering
\caption{System layouts used for simulations.}
\label{tab:syslayouts}
\begin{tabular}{|c|c|c|c|c|}
\hline
 & Scenario & $d_x$ & $d_y$ & $d_\mathrm{rb}$ \\ \hline
 Layout A & Balanced direct and RIS links & 27 m & 5 m & 40 m\\ \hline
 Layout B & Dominant direct link & 20 m & 5 m & 40 m\\ \hline
 Layout C & Dominant RIS link & 35 m & 5 m & 40 m\\ \hline
 Layout B & Stronger, non-dominant RIS link & 29 m & 5 m & 40 m\\ \hline
\end{tabular}
\end{table}

The steering vectors $\mathbf{a}_\mathrm{b}$ and $\mathbf{a}_\mathrm{r}$ correspond to the vertical uniform rectangular array (VURA) model in  \cite{miller_analytical_2019}. We consider both the sinc correlation model, which is detailed in \cite{bjornson_rayleigh_2021} and commonly applied to RIS systems, and the exponential decay model, which assumes correlation decreases exponentially with element separation \cite{loyka_channel_2001}. The sinc correlation model states that
\begin{equation}
    \mathbf{R}_{n,m} = \mathrm{sinc} \left( 2 \,d_{m,n} \right),\quad n,m = 1, \dots, L, \label{eq:corrsinc}
\end{equation}
where $\mathbf{R}\in\{\mathbf{R}_\mathrm{d},\mathbf{R}_\mathrm{ur}\}$, $d_{m,n}$ is the Euclidean distance between BS antennas/RIS elements $m$ and $n$, measured in wavelength units, and $L$ is the number of BS antennas/RIS elements. The correlation values given by (\ref{eq:corrsinc}) for realistic distances tend to result in low correlations. Therefore, to investigate system behaviour over the full range of correlation values, we also use the exponential correlation model, where
\begin{equation}
	\mathbf{R}_{n,m} = \rho^\frac{d_{m,n}}{d},
    \label{eq:exp_corr}
\end{equation}
and $0 {\leq} \rho {\leq} 1$ is the correlation level, with $\rho{=}0$ and $\rho{=}1$ giving an uncorrelated and fully correlated channel, respectively. Note that while all analytical results are valid for arbitrary RIS element spacings, \(0.1\lambda\) is used in simulations to model a dense RIS and more clearly highlight the effects of spatial correlation.

We use the classical temporal correlation models $\rho_\mathrm{d}(\tau) = J_0(2\pi f_\mathrm{d}\tau)$ for the direct channel and $\rho_\mathrm{ur}(\tau) = J_0(2\pi f_\mathrm{ur}\tau)$ for the RIS-UE channel \cite{jakes_microwave_1974}, where $f_\mathrm{d}$ and $f_\mathrm{ur}$ are the DFs of the UE with respect to the BS and RIS, respectively, and $\tau$ is the time difference.

The effects of UE speed can be seen in the LCR (and by extension, the AFD) values as they are normalised by the DF in the usual way \cite{ivanis_level_2008, beaulieu_level_2003}, where $f_m$ represents the relevant DF. Spatial channel correlation is unaffected by DF. However, the SNR correlation and channel ageing results vary with DF. The DF for a given scenario is
\begin{equation}
	f_m = \frac{vf_c}{c}\cos(\theta), \label{eq:dop_freq}
\end{equation}
where $v$ is the speed of the UE in m/s, $f_c$ is the carrier frequency, $\theta$ is the angle of receiver movement with respect to the transmitter and $c$ is the speed of light in a vacuum. Standard values for key parameters that are constant though all numerical results except where explicitly stated otherwise are defined in Table \ref{tab:simparams}.
\vspace{-0.5em}
\begin{table}[ht!]
\centering
\caption{Key simulation parameters.}
\label{tab:simparams}
\begin{tabular}{|c|c|c|}
\hline
 Parameter & Notation & Value \\ \hline
 Number of BS elements & $M$ & 32 \\ \hline
 Number of BS elements per row & $M_x$ & 8 \\ \hline
 Number of BS elements per column & $M_z$ & 4 \\ \hline 
 Number of RIS elements & $N$ & 128 \\ \hline
 Number of RIS elements per row & $N_x$ & 16 \\ \hline
 Number of RIS elements per column & $N_z$ & 8 \\ \hline
 Distance between BS elements & $d_\mathrm{b}$ & 0.5$\lambda$ \\ \hline
 Distance between RIS elements & $d_\mathrm{r}$ & 0.1$\lambda$ \\ \hline
 Height of the BS & $h_\mathrm{b}$ & 15 m \\ \hline
 Height of the RIS & $h_\mathrm{r}$ & 15 m \\ \hline
 Height of the UEs & $h_\mathrm{u}$ & 1.5 m \\ \hline
 Azimuth angle of departure at the RIS & $\phi_\mathrm{D}$ & $\tfrac{5\pi}{4}$ \\ \hline
 Elevation angle of departure at the RIS & $\theta_\mathrm{D}$ & $\tfrac{\pi}{2}$ \\ \hline
 Azimuth angle of arrival at the BS & $\phi_\mathrm{A}$ & $\tfrac{\pi}{4}$ \\ \hline
 Elevation angle of arrival at the BS & $\phi_\mathrm{D}$ & $\tfrac{\pi}{2}$ \\ \hline
 Path loss exponent for the direct channel & $\alpha_\mathrm{d}$ & 3.5 \\ \hline
 Path loss exponent for the RIS-BS channel & $\alpha_\mathrm{rb}$ & 2 \\ \hline
 Path loss exponent for the UE-RIS channel & $\alpha_\mathrm{ur}$ & 2.8 \\ \hline
 Carrier frequency & $f_{c}$ & 2.1 GHz \\ \hline
 DF of the UE with respect to the BS & $f_\mathrm{d}$ & 10 Hz \\ \hline
 DF of the UE with respect to the RIS & $f_\mathrm{ur}$ & 5 Hz \\ \hline
 Ricean K-factor of the RIS-BS link & $\kappa_\mathrm{rb}$ & 1 \\ \hline
\end{tabular}
\end{table}
\vspace{-1.5em}

\subsection{Level Crossing Rate}
\begin{figure}[!t]
    \centering
    \subfloat[Dominant direct link (Layout B).]{%
        \includegraphics[trim={0.9cm 0.13cm 0.9cm 0.11cm},clip,scale=0.4]{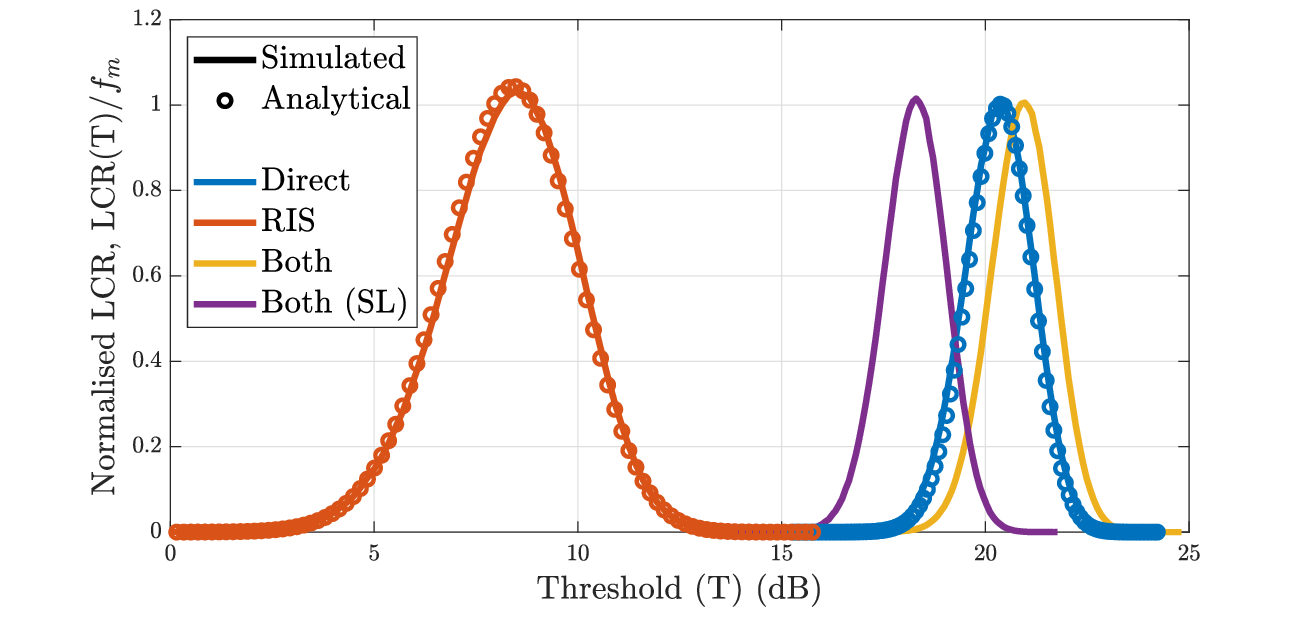}%
        \label{fig:power_variationa}
    }
    \hfil
    \subfloat[Dominant RIS link (Layout C).]{%
        \includegraphics[trim={0.9cm 0.13cm 0.9cm 0.11cm},clip,scale=0.4]{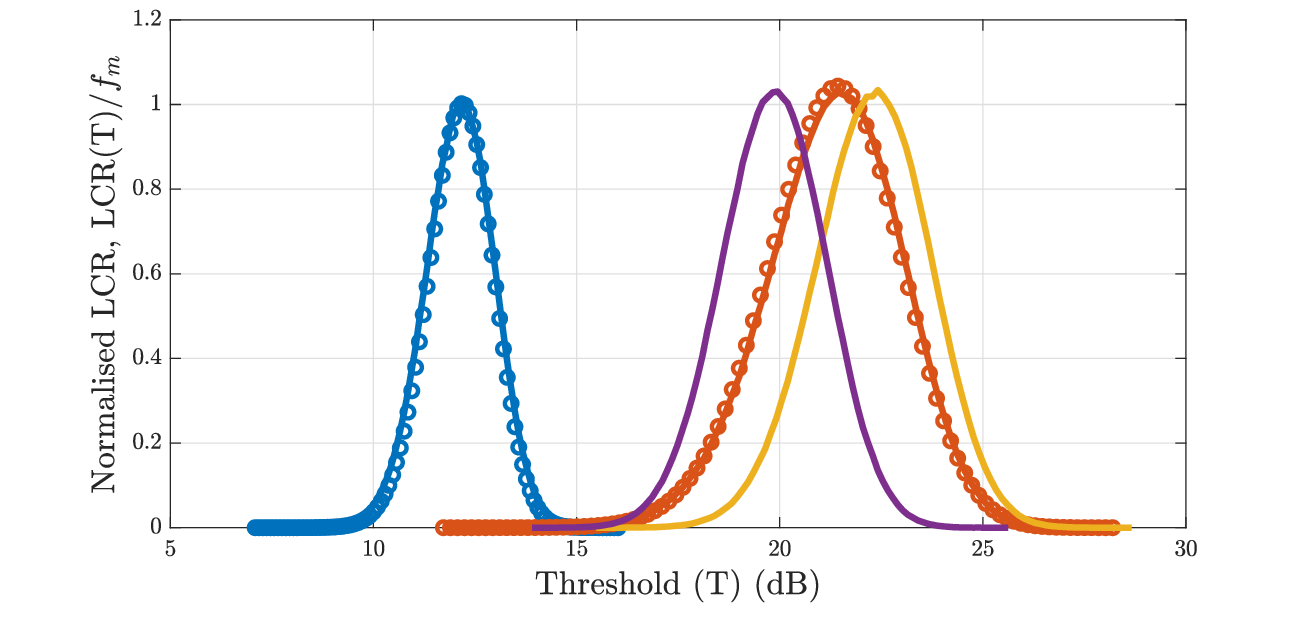}%
        \label{fig:power_variationb}
    }
    \caption{Full RIS system shown in Fig.~\ref{fig:system_diagram} with and without shadowing applied to the dominant link.}
    \label{fig:power}
\end{figure}

Figure \ref{fig:power} investigates the efficacy of the analytical methods in Sec. \ref{subsec:LCR} and the impact of the power of each link in a full channel (both direct and RIS links operational) scenario. Fig. \ref{fig:power_variationa} shows LCRs for layout B (dominant direct link) and \ref{fig:power_variationb} shows LCRs for layout C (dominant RIS link). The direct-only, RIS-only and full-channel LCRs are plotted, as well as the full channel LCR with a 50\% power loss in the dominant link (denoted shadowed link (SL)). The SL case explores the change in behaviour when the dominant link becomes less dominant.

It can be seen that the analytical LCR expressions perform well. In both Figs. \ref{fig:power_variationa} and \ref{fig:power_variationb}, they accurately match the simulated curves. This is particularly notable for the direct-only approximate LCR expression, as only the 2 leading eigenvalues were kept, due to being numerically stable and the smallest number where the CDF of the approximation was visibly indistinguishable from the original. This shows that accuracy is maintained, even while averaging large numbers of trailing eigenvalues.

When the direct and RIS links are combined, the resulting LCR resembles the LCR of the dominant link, with a slight gain in SNR provided by the weaker link shifting the LCR to the right. Reducing the power by 50\% in the dominant link resulted in an approximate 3 dB shift in the LCR curve, but minimal change in shape, showing that the temporal behaviour is largely determined by the dominant link. 

\subsection{Average Fade Duration}
\begin{figure}[ht]
    \centering
    \includegraphics[scale=0.51]{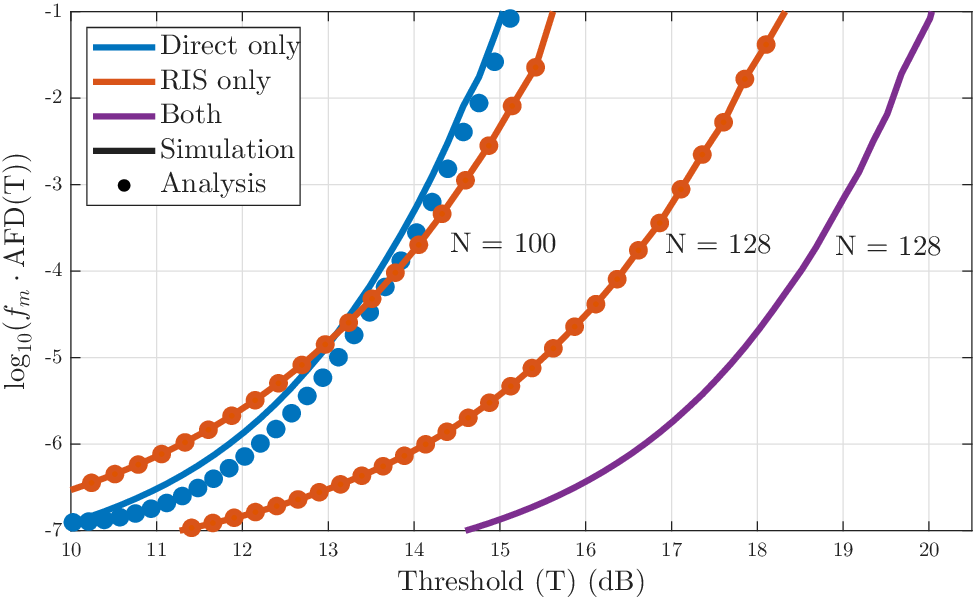}
    \caption{AFDs for the RIS system in Fig. \ref{fig:system_diagram} with layout D.}
    \label{fig:AFDs}
\end{figure}

Figure \ref{fig:AFDs} compares the approximate analytical AFDs with simulations for layout D, where the RIS-link is slightly stronger than the direct-only link. The direct-only, RIS-only and full channel AFDs are plotted for the $N=128, N_x = 16$ case. The RIS only link is also plotted for the $N=100, N_x = 10$ case, to show that the direct-only and RIS-only links can be made to perform similarly by varying element numbers. All AFDs are plotted on a logarithmic scale, and normalised by $f_m$, the relevant DF.

These results validate the performance of the analytical AFD approximations for the direct-only and RIS-only links in Sec. \ref{subsec:AFD}. It can be seen that the RIS-only AFD approximation is indistinguishable from the original function, showing that the gamma approximation to the density function is very accurate. The direct-only AFD approximation also performs well. As the y-axis is a logarithmic scale, the errors are of the order of around $10^{-5}$. These small errors are due to the averaging of the small tail eigenvalues in the direct-only LCR. In this scenario, only the two leading eigenvalues were kept, while the rest were averaged. This is the lowest number of eigenvalues that can be kept, making the performance here a lower bound on the method's accuracy. 

Layout D involves a stronger RIS path than the direct path, but both contribute significantly to the full channel. In Fig. \ref{fig:AFDs}, the threshold for fading for the $N=128$ case increases as the paths become stronger. The weakest direct-only link fades at the lowest threshold, followed by the RIS-only link and finally the strongest full channel. When the number of RIS elements is decreased to $N=100$, the RIS-only link fades at a similar threshold to the direct-only link. This shows that increasing the system size or adding an additional channel increases the threshold for fading, so a stronger SNR shifts curves to the right.

In \cite{inwood_level_2024}, we found that lower correlation leads to steeper LCR curves, as more averaging of the sums of RVs in (\ref{eq:SNRdir}) and (\ref{eq:RISonlySNR}) occurs. The distance between antennas is larger at the BS than at the RIS. Therefore, in Fig. \ref{fig:AFDs}, the lower correlation at the BS results in a steeper direct-only AFD curve.

\subsection{SNR Correlation}

\subsubsection{Variation of Link Powers}
\begin{figure}[ht]
    \centering
    \includegraphics[scale=0.51]{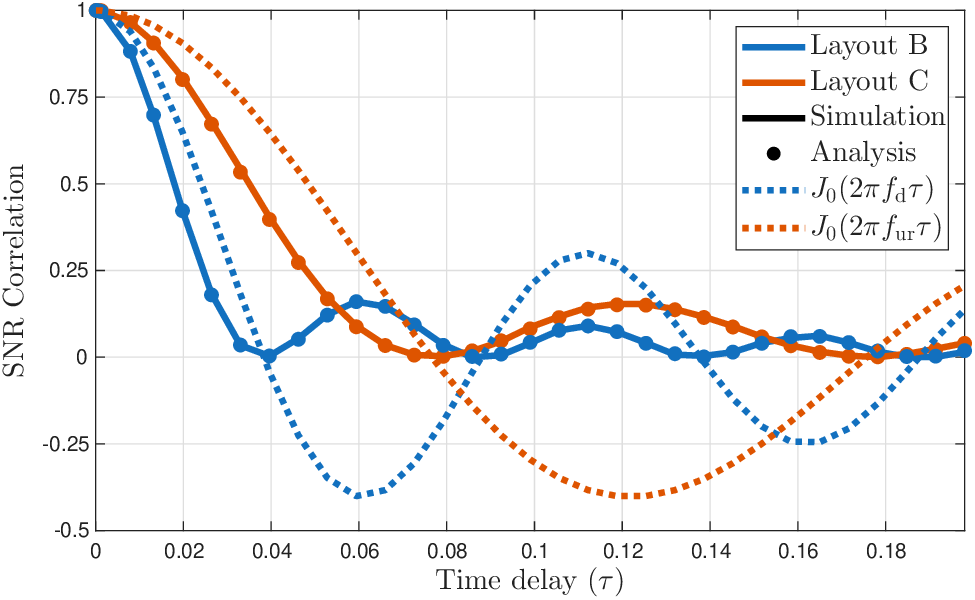}
    \caption{SNR correlation for a system with a dominant direct link (layout B) and a dominant RIS link (layout C) compared with their temporal correlation patterns.}
    \label{fig:SNRcorr_bessel}
\end{figure}

Fig. \ref{fig:SNRcorr_bessel} considers the SNR correlation for layout B (a dominant direct link) and layout C (a dominant RIS link). The Bessel functions that represent the temporal correlations at time $\tau$ are also plotted for comparison. The DF of the UE with respect to the BS assumes the UE is moving directly away from the BS ($\theta = 0$) at the average walking speed of $1.42$ m/s. The DF of the UE with respect to the RIS, assuming the same walking speed and carrier frequency, results in the UE moving at an angle of $\theta = \frac{\pi}{3}$ away from the RIS. 

It can be seen from Fig. \ref{fig:SNRcorr_bessel} that, for layouts B and C, the SNR correlation follows the correlation pattern of the dominant link, $J_0(2\pi f_m \tau)$. The maxima of the SNR correlation function and the extrema of the correlation pattern occur at the same time instant. As discussed in Remark \ref{rem:SNRcorr}, the SNR correlation is never negative as the SNR is essentially quadratic ($\mathrm{SNR}=\frac{E_s}{\sigma^2}||\mathbf{h}||^2$), and so negative correlations are transformed into positive. This leads to zero-crossings and troughs in the Bessel function correlation pattern corresponding to local minima and peaks in the SNR correlation, respectively. This indicates that any temporal correlation is helpful, including negatively correlated channels. Also, higher DFs lead to a faster decrease in SNR correlation. This is due to $\mathrm{SNR}(t)$ changing more rapidly when the UE is moving quickly, or the carrier frequency is higher.

\subsubsection{Impact of Doppler Frequency}
\begin{figure}[ht]
    \centering
    \includegraphics[scale=0.51]{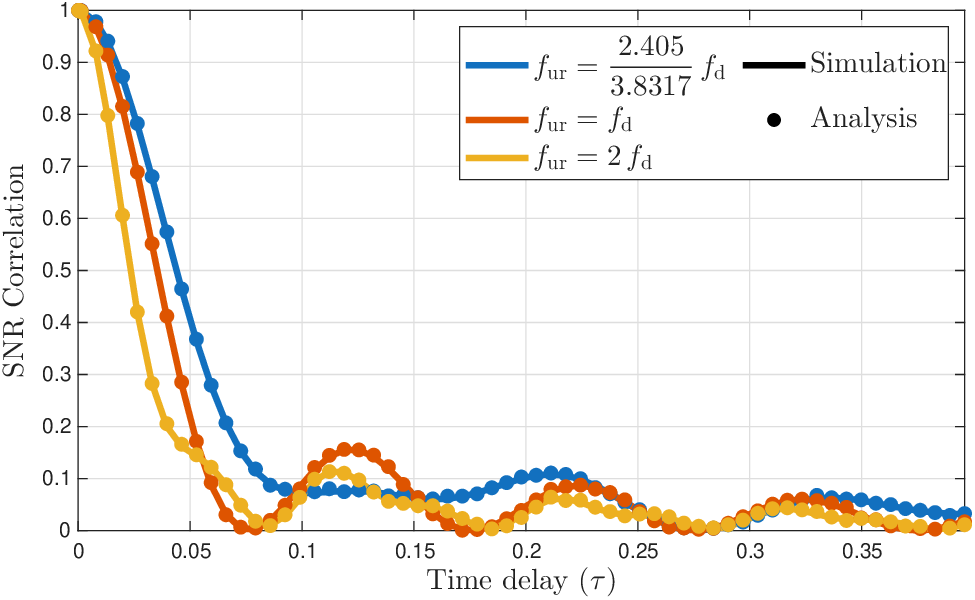}
    \caption{SNR correlation for three combinations of DFs of the UE with respect to the BS and RIS using layout A.}
    \label{fig:SNRcorr_multi}
\end{figure}

Figure \ref{fig:SNRcorr_multi} investigates the impact of DF combinations on SNR correlation for a system with equal power in the direct and RIS links (layout A). For these results, $f_\mathrm{ur}$ is varied as a simulation parameter. Results for $f_\mathrm{ur} = \{f_\mathrm{d},\, 2f_\mathrm{d},\, \frac{2.4048}{3.8317}f_\mathrm{d}\}$ are plotted. Assuming carrier frequency and UE walking speed are held constant, this corresponds to movement at angles of $\theta=\{\frac{\pi}{3},0,1.5660\}$ with respect to the RIS. $f_\mathrm{ur}=\frac{2.4048}{3.8317}f_\mathrm{d}$ is selected to achieve some cancellation of the temporal correlation of direct and RIS links. The first zero crossing of $J_0(x)$ occurs at $x = 2.4048$, and the first trough occurs at $x=3.8317$. Therefore, $J_0(\frac{2.4048}{3.8317}x)$ is zero valued when $J_0(x)$ is at its global minima, which, as SNR correlation is always positive, corresponds to a maxima of SNR correlation. 

From Fig. \ref{fig:SNRcorr_multi}, it can be seen that the SNR correlation is a mixture of oscillations when the link is balanced and $f_\mathrm{d} \neq f_\mathrm{ur}$. The impacts of each link can add constructively or destructively, depending on the specific frequencies. As evidenced when $f_\mathrm{ur} = \frac{2.4048}{3.8317}\,f_\mathrm{d}$, oscillation can be significantly reduced by opposing frequencies. However, when $f_\mathrm{d} = f_\mathrm{ur}$, the curve follows the Bessel function as in Fig. \ref{fig:SNRcorr_bessel}, where the zeros and minima of the Bessel function correspond to the zeros and maxima respectively of the SNR correlation. Also, the rate of decay at low $\tau$ is governed by both DFs. When $f_\mathrm{ur} + f_\mathrm{d}$ is larger, the curve decays faster than when it is smaller. Therefore, the behaviour of the SNR correlation is heavily dependent on both DFs.

\subsection{Channel Correlation}
We investigate channel correlation in two ways: through summing the amplitudes of all elements and considering the leading eigenvalues of the correlation matrix as in \cite{neil_on_2018}.

\subsubsection{Sum of Channel Correlation Elements}

\begin{figure}[ht]
    \centering
    \includegraphics[scale=0.51]{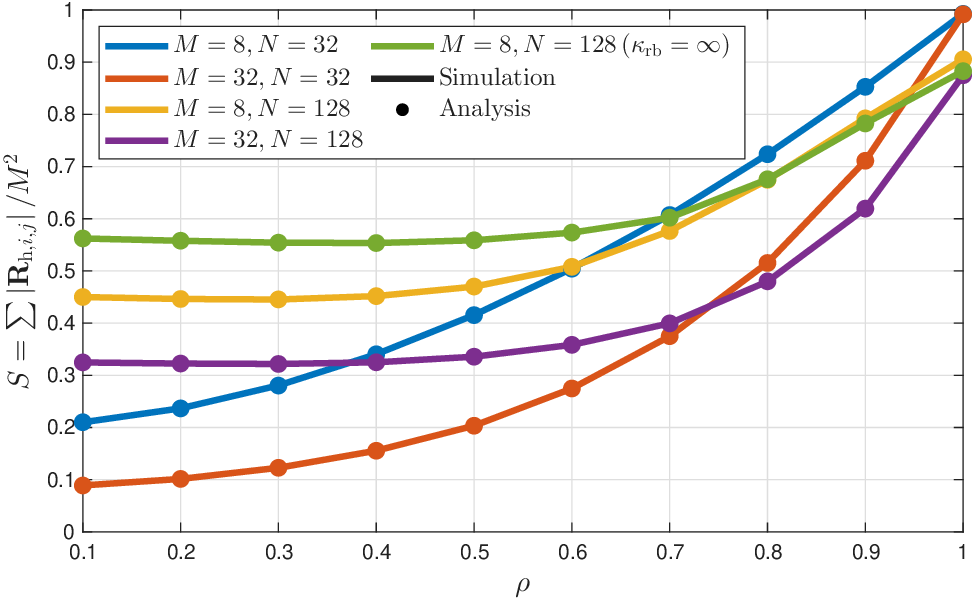}
    \caption{The normalised sum of channel correlation elements for a range of $\rho$. $\kappa_\mathrm{rb} = 1$ except where otherwise specified.}
    \label{fig:channelcorr}
\end{figure}

Let $S$ be the sum of the absolute values of all channel correlation elements normalised by the largest possible value ($M^2$ if all elements are fully correlated). Fig. \ref{fig:channelcorr} plots $S$ for a range of $\rho$ values. Notably, the exponential decay spatial correlation model in (\ref{eq:exp_corr}) is used for these results rather than the sinc spatial correlation model, to investigate a wider range of correlation values.

It is clear from Fig. \ref{fig:channelcorr} that $S$ monotonically increases with $\rho$. However, even as $\rho$ approaches 0 and correlation decreases in the scattered components of the link, $S$ does not approach $\frac{1}{M^2}$. This occurs when $\kappa_\mathrm{rb} \neq 0$, as while there is no contribution to the channel correlation from the scattered components, the LoS components still contribute. As discussed in Remark \ref{rem:uncorr}, the covariance matrix never becomes diagonal, and there will always be off diagonal covariances that will not be removed. Similarly, as discussed in Remark \ref{rem:corr}, when $\kappa_\mathrm{rb} \neq 0$, channel correlation is $< 1$ when $\rho = 1$. This is due to the different phase shifts of the LoS components. Because the other terms in \(\mathbf{D}\) come from random fading channels with varying magnitudes and phases, they do not align perfectly with the LoS component, whose entries have magnitude one but generally varying phases. Only if all phases of \(\mathbf{a}_\mathrm{b}\) were identical (an ideal case occurring at broadside for phased arrays) could $S=1$.

When $M$ is increased, $S$ decreases due to the increased size of the receive array. The average correlation between elements decreases as the distance between them increases. For example, a receive array where $M= 32$ leads to gaps four times as large being included in the average when compared to the $M=8$ case. When $N$ is increased, the size and dominance of the RIS link increases, including the LoS component. Therefore, the greater impact from $\mathbf{H}_\mathrm{rb}^{\mathrm{LoS}}$ leads to a higher $S$ at low $\rho$ and a lower $S$ at $\rho = 1$. Increasing $\kappa_\mathrm{rb}$ also increases the impact of $\mathbf{H}_\mathrm{rb}^{\mathrm{LoS}}$, and results in similar behaviour as increasing $N$. 

\subsubsection{Channel Correlation Matrix Eigenvalues}

\begin{figure}[htbp]
    \centering
    \subfloat[$M = 8$, $N = 64$, $\kappa_\mathrm{rb}=1$]{%
    \includegraphics[width=0.23\textwidth]{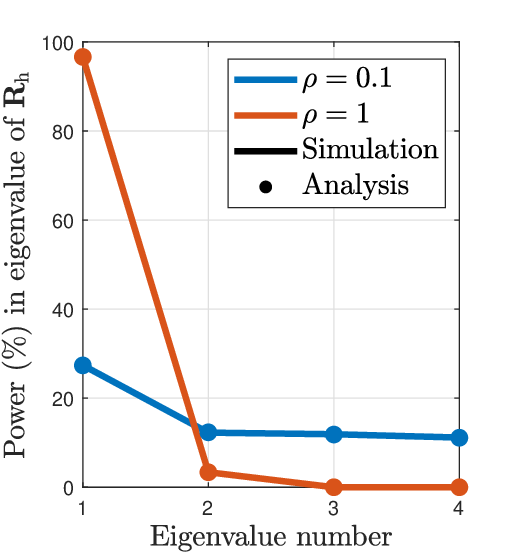}%
    \label{fig:eigcorr_a}
    }
    \hfill
    \subfloat[$M = 32$, $N = 128$, $\kappa_\mathrm{rb} = \infty$]{%
        \includegraphics[width=0.23\textwidth]{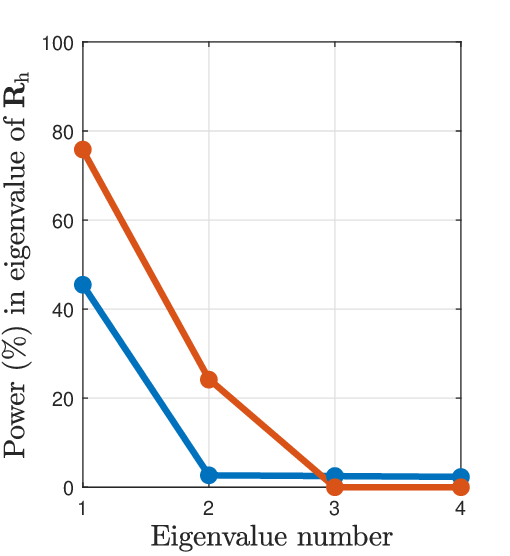}%
        \label{fig:eigcorr_d}
    }
    \caption{The percentage of the sum of all channel correlation matrix eigenvalues accounted for by the four leading eigenvalues.}
    \label{fig:eigcorrgrid}
\end{figure}

Figure \ref{fig:eigcorrgrid} considers the impact of $N$, $M$ and $\kappa_\mathrm{rb}$ on the leading eigenvalues of the channel correlation matrix, as in \cite{neil_on_2018}. As the eigenvalues plateau from number 2 or 3 for this particular system, we have plotted the first four in each plot. Again, the exponential decay correlation model is used, and results are shown for $\rho = 0.1$, a low correlation case, and $\rho = 1$, the fully correlated case.
\vspace{-1em}
When $\rho = 1$, the channel correlation matrix is rank-2. In this case, all elements of correlated Rayleigh channel vectors and matrices have the same value. Therefore, the channel vector, $\mathbf{h}$, can be simplified to a sum of two rank-1 vectors,
\begin{equation}
	\mathbf{h} \!=\! \mathbf{h}_{\mathrm{d},1}\mathbf{1}
	\!+\! c_1\mathbf{a}_\mathrm{b}\mathbf{a}_\mathrm{r}^\dagger\mathbf{\Phi}\mathbf{h}_\mathrm{ur}\!+\! \mathbf{G}_{1,1}\mathbf{11}^T
	\mathbf{\Phi}\mathbf{h}_\mathrm{ur} \!=\! d_1\mathbf{1} \!+\!d_2\mathbf{a}_\mathrm{b},
\end{equation}
where $\mathbf{1} = [1 \dots 1]^T$, $d_1 = \mathbf{h}_{\mathrm{d},1} + \mathbf{G}_{1,1}\mathbf{1}^T\mathbf{\Phi h}_\mathrm{ur}$, $d_2 = c_1\mathbf{a}_\mathrm{r}^\dagger\mathbf{\Phi}\mathbf{h}_\mathrm{ur}$, $\mathbf{h}_{\mathrm{d},1}$ is the first element in the constant vector $\mathbf{h}_{\mathrm{d}}$, and $\mathbf{G}_{1,1}$ is the first element in the constant matrix $\mathbf{G}$. The correlation matrix is thus
\begin{align}
	\mathbb{E}\left[\mathbf{h}\mathbf{h}^\dagger\right] =& \mathbb{E}\left[\Big(d_1\mathbf{1} + d_2\mathbf{a}_\mathrm{b}\Big)\left(d_1^*\mathbf{1}^T + d_2^*\mathbf{a}_\mathrm{b}^\dagger\right)\right], \notag \\
	 =& \mathbf{1}\big(\mathbb{E}\left[|d_1|^2\right]\mathbf{1}^T + \mathbb{E}\left[d_1d_2^*\right]\mathbf{a}_\mathrm{b}^\dagger \big) \notag \\ &\qquad\quad+ \mathbf{a}_\mathrm{b}\big(\mathbb{E}\left[|d_2|^2\right]\mathbf{a}_\mathrm{b}^\dagger + \mathbb{E}\left[ d_1^*d_2\right]\mathbf{1}^T \big).
\end{align}
Both of these components are rank-1. Therefore, when $\rho = 1$, the two independent rank-1 components result in an overall rank-2 correlation matrix.

While eigenvalues correspond to a mixture of the LoS and scattered components, one eigenvalue mainly corresponds to each of the two rank-1 correlation matrices. Whether the eigenvalue dominated by the  LoS component is larger depends on a range of factors, including K-factor, link correlation, $M$, $N$ and link power. For the parameters used in these results, when $\rho = 0.1$, the LoS dominated eigenvalue is larger, as the scattered link is almost uncorrelated. However, when $\rho=1$, the scattered dominated eigenvalue is larger, due to the scattered channel component being perfectly correlated and the LoS channel component never reaching 1 due to the different phases of each element of $\mathbf{a}_\mathrm{b}$ (as explained above). 

Increasing the K-factor reduces the gap between the scattered and LoS dominated eigenvalues when $\rho = 1$, as seen in Fig. \ref{fig:eigcorr_d} when compared to Fig. \ref{fig:eigcorr_a}. Increasing K-factor increases the impact of the LoS component, as less spatial multiplexing is available. Increasing $N$ has a similar impact, as this increases the effects of the RIS link. Only the RIS link contributes to the LoS component, which results in a greater percentage of power in the corresponding eigenvalue.

Finally, increasing $M$ reduces the value of the trailing eigenvalues when $\rho = 0.1$ as the total power is spread across more channels. In Fig. \ref{fig:eigcorr_a} where $M = 8$, eigenvalues 3 and above have approximately 10\% of the total power each, while in Fig. \ref{fig:eigcorr_d} where $M=32$, eigenvalues 3 and above have only approximately 4\% of the total power.

\subsection{Channel Ageing Loss}

\begin{figure}[ht]
    \centering
    \includegraphics[scale=0.51]{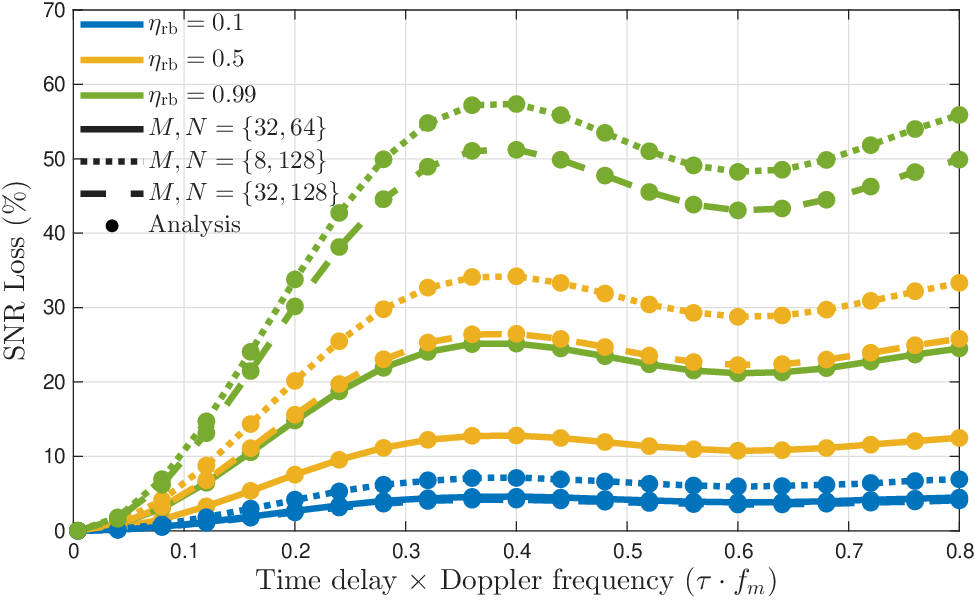}
    \caption{The SNR loss occurring for channels at time $t+\tau$ with an RIS designed at time $t$.}
    \label{fig:ageingloss}
\end{figure}

Figure \ref{fig:ageingloss} shows the SNR loss that occurs at time $t+\tau$ when the RIS design has not been updated since time $t$ for a range of $\eta_\mathrm{rb}$ and system dimensions. Defined in (\ref{eq:etarb}), $\eta_\mathrm{rb}$ represents the proportion of the BS-RIS that is LoS. The SNR loss is given as a percentage of the initial SNR at time $t=0$. The time delay values are normalised by $f_m$, the relevant DF (where $f_m = f_\mathrm{d} = f_\mathrm{ur}$). Comparing the SNR loss to a time-frequency product means that results are consistent and comparable across DFs. The sinc correlation model is used for these results.

For fixed array dimensions, as the LoS percentage of $\mathbf{H}_\mathrm{rb}(t)$ increases, the percentage SNR loss due to ageing also increases. This occurs because the RIS phase matrix $\mathbf{\Phi}(t)$ is designed to align the LoS component of $\mathbf{H}_\mathrm{rb}$ with the instantaneous UE-RIS channel $\mathbf{h}_\mathrm{ur}(t)$, which is time-varying. When $\mathbf{H}_\mathrm{rb}$ is strongly LoS, this alignment is initially more effective, resulting in a higher SNR at the time of phase selection. However, as $\mathbf{h}_\mathrm{ur}(t)$ evolves due to ageing, the mismatch between the phase design and the current channel increases. As the initial SNR is higher in the strongly LoS case, the relative degradation is more pronounced, leading to a greater percentage SNR loss. In contrast, when $\mathbf{H}_\mathrm{rb}$ is less dominated by LoS, the initial phase design is less effective, and therefore the ageing-induced loss appears smaller in percentage terms. Once the channel has aged enough that the loss has reached a peak, the SNR loss follows the temporal correlation structure, which in this case is the zero-th order Bessel function. $\mathrm{SNR}(t)$ and $\mathrm{SNR}(t+\tau)$ move between being positively and negatively correlated, leading to oscillations in SNR loss.

In 5G standards, it is stated that resource blocks can range from 62.5 $\mu$s to 1 ms, depending on subcarrier spacing \cite{etsi_5g_2019}. Therefore, assuming a DF of 132 Hz as in the previous section, the maximum time-frequency product for one resource block is 0.132. Therefore, the maximum SNR loss after one resource block is around $15\%$ when $M =32$, $N=128$ and $\eta_\mathrm{rb}=0.99$, far below the maximum percentage loss of $51\%$. This would be even lower for shorter resource blocks and lower DFs.

For larger $\eta_\mathrm{rb}$, a greater percentage of SNR is lost due to ageing when $N$ is significantly larger than $M$. When $\eta_{\mathrm{rb}} = 0.99$ and $\eta_{\mathrm{rb}} = 0.5$, the most loss occurs when $M,N = \{8, 128\},$ followed by $M,N = \{32, 128\}$. Significantly lower losses occur when $M,N = \{32, 64\}$. When the RIS link is dominant, which occurs when the LoS component is more dominant (a high K-factor) or when $N$ is significantly larger than $M$, the RIS design is more influential to the SNR. Therefore, the SNR loss is higher in this scenario. As discussed, when $\eta_\mathrm{rb}$ is smaller, the design is less accurate so the RIS design ageing has less effect on the SNR. Therefore, the difference between the $M,N = \{32, 128\}$ and $M,N = \{32, 64\}$ curves is much less when $\eta_\mathrm{rb}=0.1$. 

\section{Conclusion}
We investigated the second order channel statistics of an SU RIS system with RIS element reflection coefficients set according to the method proposed in \cite{singh_optimal_2021}. Assuming an LoS RIS-BS link, we stated the exact expression for the LCR of the RIS only channel, and a numerically stable approximation to the LCR of MRC in correlated Rayleigh fading with large numbers of antennas. The RIS-only LCR has the same form as the LCR of an uncorrelated MRC system, showing that RISs do not amplify temporal changes in the channel. These expressions were then used to find expressions for the AFD of the RIS only and direct only links. We also derived exact expressions for the correlation of the SNR at time $t$ and the SNR at time $t+\tau$. For a Ricean RIS-BS link, we derived exact expressions for the spatial channel correlation matrix and the SNR at time $t + \tau$, the latter of which allowed us to find the SNR loss between times $t$ and $t+\tau$.

We showed that increasing the power in a link leads to the overall system SNR LCR becoming more similar to the LCR of that link. Also, increasing the number of antenna elements leads to a steeper AFD curve. It was found that when either the direct or RIS link is dominant, the SNR correlation follows the temporal correlation of the dominant link, which is dependent on DF. However, the SNR correlation is always positive due to the SNR being quadratic, indicating that any temporal correlation, positive or negative, is helpful.

We investigated the normalised sum of elements and the percentage of power in the leading eigenvalues of the spatial channel correlation matrix. As our model makes the reasonable assumption that the RIS-BS link is Ricean, the line of sight component leads to the channel covariance matrix never becoming diagonal when the Rayleigh channel components are uncorrelated, or a matrix of ones when the Rayleigh channel components are perfectly correlated. This is exacerbated when $N$ increases and the LoS component becomes more influential. When the Rayleigh channel components are fully correlated, the leading eigenvalues indicate the channel correlation matrix is rank-2, as opposed to full rank when correlation is not equal to 1. This is due to the overall correlation matrix collapsing to an LoS dominated rank-1 component and a scattered dominated rank-1 component.

Finally, it was seen that the SNR ageing loss as a percentage was highest when the RIS-BS link had a stronger LoS component. The phase selection method is designed for this scenario, and as the SNR at $t=0$ is higher, the ageing of the RIS design leads to greater losses. Also, the losses are greatest when $N$ is significantly larger than $M$, as the RIS design influences a greater proportion of the SNR.

\section*{Appendix A \\ Derivation of $\epsilon$}
We compute $\epsilon$ by taking the expectation of the six terms in (\ref{eq:SNR_correlation}), which is done below. \\ 

\textit{Term 1}: $\mathbb{E}[a(t)a(t+\tau)]$ \\
Expanding out $\mathbb{E}[a(t)a(t+\tau)]$ and using the temporal correlation to write $\mathbf{h}_\mathrm{d}(t+\tau)$ in terms of $\mathbf{h}_\mathrm{d}(t)$,
\begin{multline}
    \mathbb{E}[a(t)a(t+\tau)]\!=\!\mathbb{E}\Big[\mathbf{h}_\mathrm{d}^\dagger(t)\mathbf{h}_\mathrm{d}(t)\!\Big(\!\rho_\mathrm{d}^*(\tau)\mathbf{h}_\mathrm{d}^\dagger(t) + \sqrt{\!1\!-\!|\rho_\mathrm{d}(\tau)|^2} \\  \times\mathbf{e}_\mathrm{d}^\dagger(t)\Big)\Big(\rho_\mathrm{d}(\tau)\mathbf{h}_\mathrm{d}(t) + \sqrt{1\!-\!|\rho_\mathrm{d}(\tau)|^2}\mathbf{e}_\mathrm{d}(t)\Big)\Big], \notag 
\end{multline}
where $\mathbf{e}_\mathrm{d}(t)\sim \mathcal{CN}(0,\beta_\mathrm{d}\mathbf{R}_\mathrm{d})$. Dropping time notation as all are at time $t$, and using the independence  of $\mathbf{h}_\mathrm{d}$ and $\mathbf{e}_\mathrm{d}$,
\begin{multline}
    \mathbb{E}[a(t)a(t+\tau)]\!=\!|\rho_\mathrm{d}(\tau)|^2\mathbb{E}\Big[\big(\mathbf{h}_\mathrm{d}^\dagger\mathbf{h}_\mathrm{d}\big){}^2\Big] \\ + \big(1-|\rho_\mathrm{d}(\tau)|^2\big)
    \mathbb{E}\big[\mathbf{h}_\mathrm{d}^\dagger\mathbf{h}_\mathrm{d}\big]\mathbb{E}\big[\mathbf{e}_\mathrm{d}^\dagger\mathbf{e}_\mathrm{d}\big]. \label{eq:Eatatau_appen}
\end{multline}
Combining the known results from \cite{singh_optimal_2021} for $\mathbb{E}\big[\mathbf{h}_\mathrm{d}^\dagger\mathbf{h}_\mathrm{d}\big]$ and $\mathbb{E}\big[\mathbf{e}_\mathrm{d}^\dagger\mathbf{e}_\mathrm{d}\big]$ with (\ref{eq:Eatatau_appen}) gives (\ref{eq:Eatatau}).\\

\textit{Term 2}: $\mathbb{E}[a(t)b(t+\tau)]$ \\
Expanding and using the independence of $Y$ and $\mathbf{h}_\mathrm{d}$,
\begin{equation}
    \mathbb{E}[a(t)b(t+\tau)]\!=\!\!\sqrt{\!\beta_\mathrm{rb}} \mathbb{E}\!\big[Y\!(t+\tau)\big]\mathbb{E}\!\!\left[\mathbf{h}_\mathrm{d}^\dagger(t)\mathbf{h}_\mathrm{d}(t)|\mathbf{a}_\mathrm{b}^\dagger\mathbf{h}_\mathrm{d}(t\!+\!\tau)|\right]\!\!. \notag
\end{equation}
Rewriting $\mathbf{h}_\mathrm{d}(t)$ in terms of $\mathbf{h}_\mathrm{d}(t+\tau)$ using the temporal correlation,
\begin{align}
    \mathbb{E}[a(t)b(t&+\tau)] = \sqrt{\!\beta_\mathrm{rb}}\, \mathbb{E}\big[Y(t\!+\!\tau)\big]\mathbb{E}\big[\big(\rho_\mathrm{d}^*(\tau)\mathbf{h}_\mathrm{d}^\dagger(t\!+\!\tau) \notag \\ & + \sqrt{1-|\rho_\mathrm{d}(\tau)|^2}\mathbf{e}^\dagger_\mathrm{d}(t+\tau)\big)\big(\rho_\mathrm{d}(\tau)\mathbf{h}_\mathrm{d}(t\!+\!\tau) \notag \\ & + \sqrt{1-|\rho_\mathrm{d}(\tau)|^2} \mathbf{e}_\mathrm{d}(t\!+\!\tau)\big)|\mathbf{a}_\mathrm{b}^\dagger\mathbf{h}_\mathrm{d}(t\!+\!\tau)|\big]. \notag
\end{align}
Dropping time notation as all terms are at time $t+\tau$, expanding out brackets and removing zero mean terms,
\begin{multline}
    \mathbb{E}[a(t)b(t+\tau)]\!=\!\sqrt{\!\beta_\mathrm{rb}}\, \mathbb{E}\big[Y\big]\big(|\rho_\mathrm{d}(\tau)|^2\mathbb{E}\big[\mathbf{h}_\mathrm{d}^\dagger \mathbf{h}_\mathrm{d}|\mathbf{a}_\mathrm{b}^\dagger\mathbf{h}_\mathrm{d}|\big] \\ +\big(1\!-\!|\rho_\mathrm{d}(\tau)|^2\big)\mathbb{E}\big[\mathbf{e}_\mathrm{d}^\dagger\mathbf{e}_\mathrm{d}\big]\mathbb{E}\big[|\mathbf{a}_\mathrm{b}^\dagger\mathbf{h}_\mathrm{d}|\big]\big). \label{eq:atbtau_appen}
\end{multline}
$\mathbb{E}[Y]$ is given in (\ref{eq:EY}), results from \cite{singh_optimal_2021} give $\mathbb{E}[\mathbf{e}_\mathrm{d}^\dagger\mathbf{e}_\mathrm{d}]$ and $\mathbb{E}\big[|\mathbf{a}_\mathrm{b}\mathbf{h}_\mathrm{d}|\big]$ is given in \cite[(4.2)]{miller_complex_1974}. To derive $\mathbb{E}\big[\mathbf{h}_\mathrm{d}^\dagger \mathbf{h}_\mathrm{d}|\mathbf{a}_\mathrm{b}^\dagger\mathbf{h}_\mathrm{d}|\big]$, the expression can be rewritten as
\begin{equation}
    \mathbb{E}\big[\mathbf{h}_\mathrm{d}^\dagger \mathbf{h}_\mathrm{d}|\mathbf{a}_\mathrm{b}^\dagger\mathbf{h}_\mathrm{d}|\big] = \sum_{k=1}^M\mathbb{E}\left[|\mathbf{h}_{\mathrm{d},k}|^2|\mathbf{a}_\mathrm{b}^\dagger\mathbf{h}_\mathrm{d}|\right], \label{eq:sumhd2abhd}
\end{equation}
allowing the terms that sum to $\mathbf{h}_\mathrm{d}^\dagger\mathbf{h}_\mathrm{d}$ to be considered independently. Let $X_1 = \mathbf{h}_{\mathrm{d},k}$, where ${X_1}/{\sqrt{\beta_\mathrm{d}}}\sim\mathcal{CN}(0,1)$, and $X_2 = \mathbf{h}_\mathrm{d}^\dagger\mathbf{a}_\mathrm{b}$, where ${X_2}/{\sqrt{\beta_\mathrm{d}\mathbf{a}_\mathrm{b}\mathbf{R}_\mathrm{d}\mathbf{a}_\mathrm{b}}}\sim\mathcal{CN}(0,1)$. $X_1$ can be written in terms of $X_2$ using their correlation coefficient, $\rho_{12}$, so that
\begin{equation}
    \frac{X_1}{\sqrt{\beta_\mathrm{d}}} = \rho_{12}\frac{X_2}{\sqrt{\beta_\mathrm{d}\mathbf{a}_\mathrm{b}^\dagger\mathbf{R}_\mathrm{d}\mathbf{a}_\mathrm{b}}} + \sqrt{1-|\rho_{12}|^2}\,e_{12}, \label{eq:linkingX1X2}
\end{equation}
where $e_{12}\sim\mathcal{CN}(0,1)$. To determine $\rho_{12}$, known expectations of $X_1$ and $X_2$ can be used. Multiplying both sides by $X_2$, taking the expectation and removing zero mean terms,
\begin{equation}
    \frac{\mathbb{E}[X_1X_2^*]}{\sqrt{\beta_\mathrm{d}}} = \frac{\rho_{12}\mathbb{E}[|X_2|^2]}{\sqrt{\beta_\mathrm{d}\mathbf{a}_\mathrm{b}^\dagger\mathbf{R}_\mathrm{d}\mathbf{a}_\mathrm{b}}}.
\end{equation}
From \cite[(4.2)]{miller_complex_1974}, $\mathbb{E}\left[|X_2|^2\right] = \beta_\mathrm{d}\mathbf{a}_\mathrm{b}^\dagger\mathbf{R}_\mathrm{d}\mathbf{a}_\mathrm{b}$, which gives $\rho_{12} = \mathbf{R}_{\mathrm{d},k}\mathbf{a}_\mathrm{b}/\sqrt{\mathbf{a}_\mathrm{b}^\dagger\mathbf{R}_\mathrm{d}\mathbf{a}_\mathrm{b}}$. Using these results,  rearranging (\ref{eq:linkingX1X2}) for $X_1$ and rewriting (\ref{eq:sumhd2abhd}) solely in terms of $X_2$,
\begin{equation}
\mathbb{E}\!\left[|X_1|^2|X_2|\right] = \frac{|\rho_{12}|^2\,\mathbb{E}\big[|X_2|^3\big]}{\mathbf{a}_\mathrm{b}^\dagger\mathbf{R}_\mathrm{d}\mathbf{a}_\mathrm{b}} + \beta_\mathrm{d}(1-|\rho_{12}|^2)\mathbb{E}[|X_2|]. \notag
\end{equation}
Applying \cite[(4.2)]{miller_complex_1974} again to solve $\mathbb{E}[|X_2|^3]$ gives
\begin{multline}
    \mathbb{E}\big[\mathbf{h}_\mathrm{d}^\dagger \mathbf{h}_\mathrm{d}|\mathbf{a}_\mathrm{b}^\dagger\mathbf{h}_\mathrm{d}|\big] = \frac{\sqrt{\pi}\beta_\mathrm{d}^{3/2}}{2}\bigg(\frac{3\mathbf{a}_\mathrm{b}\mathbf{R}_\mathrm{d}^2\mathbf{a}_\mathrm{b}}{2\sqrt{\mathbf{a}_\mathrm{b}\mathbf{R}_\mathrm{d}\mathbf{a}_\mathrm{b}}} \\ + \sqrt{\mathbf{a}_\mathrm{b}^\dagger\mathbf{R}_\mathrm{d}\mathbf{a}_\mathrm{b}}\left(M - \frac{\mathbf{a}_\mathrm{b}\mathbf{R}_\mathrm{d}^2\mathbf{a}_\mathrm{b}}{\mathbf{a}_\mathrm{b}\mathbf{R}_\mathrm{d}\mathbf{a}_\mathrm{b}}\right)\!\!\!\bigg).\label{eq:hdhdabhd} 
\end{multline}
Combining the above results with (\ref{eq:atbtau_appen}) gives (\ref{eq:Eatbtau}). \\

\textit{Term 3}: $\mathbb{E}[a(t)c(t+\tau)]$ \\
Expanding $\mathbb{E}[a(t)c(t+\tau)]$ and again using the independence of $\mathbf{h}_\mathrm{d}$ and $Y$,
\begin{equation}
    \mathbb{E}[a(t)c(t+\tau)] = M\beta_\mathrm{rb}\mathbb{E}\big[\mathbf{h}_\mathrm{d}^\dagger(t)\mathbf{h}_\mathrm{d}(t)\big]\mathbb{E}\big[Y^2(t+\tau)\big]. \label{eq:atctau_appen}
\end{equation}
As each independent term is at only one time instant, the time notation can be dropped. From \cite{singh_optimal_2021},
\begin{equation}
    \!\mathbb{E}\big[Y^2\big]\!=\!\beta_\mathrm{ur}\!\bigg(\!\!N\!+\!\tfrac{\pi}{4}\!\sum_{i=1}^N\sum_{j\neq i }{}_2F_1\!\!\left(\!-\tfrac{1}{2},\!-\tfrac{1}{2},\!1,\!|\mathbf{R}_{\mathrm{ur},i,j}|^2\right)\!\!\bigg)\!.\!\label{eq:Y2}
\end{equation}
Combining \cite[(20)]{singh_optimal_2021} and (\ref{eq:Y2}) with (\ref{eq:atctau_appen}) gives (\ref{eq:Eatctau}). \\

\textit{Term 4}: $\mathbb{E}[b(t)b(t+\tau)]$ \\
Expanding $\mathbb{E}[b(t)b(t+\tau)]$ and taking the expectations of the independent terms,
\begin{equation}
    \mathbb{E}[b(t)b(t+\tau)]\!=\!\beta_\mathrm{rb}\mathbb{E}[Y\!(t)Y\!(t+\tau)]\mathbb{E}[|\mathbf{a}_\mathrm{b}^\dagger\mathbf{h}_\mathrm{d}(t)||\mathbf{a}_\mathrm{b}^\dagger\mathbf{h}_\mathrm{d}(t+\tau)|]. \label{eq:Ebtbtau_appen}
\end{equation}
(4.19) of \cite{miller_complex_1974} gives the expectation of two correlated random amplitudes as
\begin{equation}
    \mathbb{E}[r_1r_2] = \tfrac{\pi}{4}\sqrt{\mathbf{R}_{1,1}\mathbf{R}_{2,2}}\,\,{}_2F_1\left(-\tfrac{1}{2}, -\tfrac{1}{2}, 1, \lambda_{12}^2\right), \label{eq:E_r1r2}
\end{equation}
where $\mathbf{R}$ is the covariance matrix for random variables $r_1$ and $r_2$ and $\lambda_{12}^2 = \frac{|\mathbf{R}_{1,2}|^2}{\mathbf{R}_{1,1}\mathbf{R}_{2,2}}$. First, let $z_1 = r_1\mathrm{e}^{\phi_1} = \mathbf{a}_\mathrm{b}^\dagger\mathbf{h}_\mathrm{d}(t)$ and $z_2 = r_2\mathrm{e}^{\phi_2} = \mathbf{a}_\mathrm{b}^\dagger\mathbf{h}_\mathrm{d}(t+\tau)$. Therefore, using the result for $\mathbb{E}[|X_2|^2]$ from the derivation of term 2 and knowing that, in this case, $\rho_\mathrm{d}(\tau)$ is real,
\begin{equation}
    \mathbf{R} \!=\!\!
    \setlength\arraycolsep{0.8pt}
        \begin{bmatrix}
            \mathbb{E}[z_1z_1^*] & \mathbb{E}[z_1z_2^*]\\
            \mathbb{E}[z_2z_1^*] & \mathbb{E}[z_2z_2^*]
        \end{bmatrix}
   \!\!=\!\!
        \begin{bmatrix}
            \beta_\mathrm{d}\mathbf{a}_\mathrm{b}^\dagger\mathbf{R}_\mathrm{d}\mathbf{a}_\mathrm{b} & \beta_\mathrm{d}\rho_\mathrm{d}(\tau)\mathbf{a}_\mathrm{b}^\dagger\mathbf{R}_\mathrm{d}\mathbf{a}_\mathrm{b}\\
            \beta_\mathrm{d}\rho_\mathrm{d}(\tau)\mathbf{a}_\mathrm{b}^\dagger\mathbf{R}_\mathrm{d}\mathbf{a}_\mathrm{b} & \beta_\mathrm{d}\mathbf{a}_\mathrm{b}^\dagger\mathbf{R}_\mathrm{d}\mathbf{a}_\mathrm{b}
        \end{bmatrix}\!\!, \notag
\end{equation}
and thus
\begin{multline}
    \mathbb{E}[|\mathbf{a}_\mathrm{b}^\dagger\mathbf{h}_\mathrm{d}(t)||\mathbf{a}_\mathrm{b}^\dagger\mathbf{h}_\mathrm{d}(t+\tau)|] \\ = \tfrac{\pi}{4}\beta_\mathrm{d}\mathbf{a}_\mathrm{b}^\dagger\mathbf{R}_\mathrm{d}\mathbf{a}_\mathrm{b}\,{}_2F_1\!\!\left(\!-\tfrac{1}{2},-\tfrac{1}{2},1, |\rho_\mathrm{d}(\tau)|^2\right). \label{eq:Eabsabhdabhd}
\end{multline}
Letting $z_1 = \mathbf{h}_{\mathrm{ur},i}(t)$ and $z_2 = \mathbf{h}_{\mathrm{ur},j}(t+\tau)$ and applying the same process gives
\begin{align}
    \mathbb{E}[Y&(t)Y(t+\tau)] = \sum_{i=1}^N\sum_{j=1}^N\mathbb{E}\left[|\mathbf{h}_{\mathrm{ur},i}(t)||\mathbf{h}_{\mathrm{ur},j}(t+\tau)|\right]. \notag \\
    =&\beta_\mathrm{ur}\tfrac{\pi}{4}\sum^N_{i=1}\sum^N_{j=1}{}_2F_1\!\left(\!-\tfrac{1}{2},-\tfrac{1}{2},1,|\rho_\mathrm{ur}(\tau)|^2|\mathbf{R}_{\mathrm{ur},i,j}|^2\right)\!. \label{eq:EabsYtYtau}
\end{align}
Combining (\ref{eq:Eabsabhdabhd}) and (\ref{eq:EabsYtYtau}) with (\ref{eq:Ebtbtau_appen}) gives (\ref{eq:Ebtbtau}). \\

\textit{Term 5}: $\mathbb{E}[b(t)c(t+\tau)]$ \\
Expanding out $b(t)$ and $c(t+\tau)$ and separating independent terms,
\begin{equation}
    \mathbb{E}[b(t)c(t+\tau)] = M\beta_\mathrm{rb}^{3/2}\mathbb{E}\big[Y\!(t)Y^2\!(t+\tau\!)\!\big]\mathbb{E}\big[|\mathbf{a}_\mathrm{b}^\dagger\mathbf{h}_\mathrm{d}(t)|\big]\!. \label{eq:Ebtctau_appen}
\end{equation}
From Section 2.3 of \cite{kotz_continuous_2000}, the gamma approximation for $Y(t)$ leads to
\begin{equation}
    \mathbb{E}\big[Y(t)Y^2(t+\tau)\big] = \alpha^3r(r+2)(4\gamma^2+r) \label{eq:EYtYtau2}
\end{equation}
where 
\begin{equation}
    \gamma = \frac{\mathrm{Cov}\left(Y(t), Y(t+\tau)\right)}{\mathrm{Var}\left(Y(t)\right)},
    \label{eq:rho_appen}
\end{equation}
\begin{equation}
    \mathrm{Cov}\left(Y(t), Y(t+\tau)\right)\! = \!\mathbb{E}[Y(t)Y(t+\tau)]\!-\!\mathbb{E}[Y(t)]^2\!. \label{eq:covYtYtau}
\end{equation}
$\mathbb{E}\big[|\mathbf{a}_\mathrm{b}\mathbf{h}_\mathrm{d}|\big]$ was stated in the derivation of term 2. Combining $\mathbb{E}[Y(t)Y(t+\tau)]$ as given in (\ref{eq:EabsYtYtau}) and $\mathbb{E}[Y(t)]$ as defined in (\ref{eq:EY}) with (\ref{eq:covYtYtau}) gives $\gamma$ as seen in (\ref{eq:rho}). $\alpha = \frac{\mathrm{Var}[Y(t)]}{2\mathbb{E}[Y(t)]}$, $r=\frac{\mathbb{E}[Y(t)]}{\alpha}$,  and $\mathrm{Var}[Y(t)]$ is stated in (\ref{eq:varY}). Therefore, combining the above results with (\ref{eq:Ebtctau_appen}) gives (\ref{eq:Ebtctau}). \\

\textit{Term 6}: $\mathbb{E}[c(t)c(t+\tau)]$ \\
Expanding out $c(t)$ and $c(t+\tau)$ leaves only 1 independent 
term,
\begin{equation}
    \mathbb{E}[c(t)c(t+\tau)] = M^2\beta_\mathrm{rb}^2\mathbb{E}[Y^2(t)Y^2(t+\tau)].
\end{equation}
Again, from Section 2.3 of \cite{kotz_continuous_2000}, the gamma approximation for $Y(t)$ leads to
\begin{equation}
    \mathbb{E}[Y^2(t)Y^2(t+\tau)] = \alpha^4 r (r+2) (r^2 + 8r\gamma^2 + 2r + 8\gamma^4 + 16\gamma^2), \notag
\end{equation}
where $\alpha$, $r$ and $\gamma$ are as defined for (\ref{eq:EYtYtau2}).

\section*{Appendix B \\ Derivation of $\mathbb{E}[\nu^*\mathbf{h}_\mathrm{d}]$ and $\mathbf{Z}$}
$\mathbb{E}[\nu^*\mathbf{h}_\mathrm{d}]$ and $\mathbf{Z}$ are the two new terms required to derive the channel correlation in Sec. \ref{sec:chancorr}. Each of the $k$ elements in $\mathbb{E}[\nu^*\mathbf{h}_\mathrm{d}]$ can be written as
\begin{equation}
    \mathbb{E}[\nu^*\mathbf{h}_{\mathrm{d},k}] = \mathbb{E}\left[\frac{\mathbf{h}_\mathrm{d}^\dagger\mathbf{a}_\mathrm{b}}{|\mathbf{h}_\mathrm{d}^\dagger\mathbf{a}_\mathrm{b}|}\mathbf{h}_{\mathrm{d},k}\right].
\end{equation}
For two correlated random variables, $X_1$ and $X_2$, related by the expression $X_1 = \rho_{12}X_2^* + \sqrt{1-|\rho_{12}|^2}e$,
\begin{equation}
    \label{eq:corrrayleighvblangle}
    \mathbb{E}\left[\frac{X_2}{|X_2|}X_1\right] = \rho_{12}\mathbb{E}[{|X_2|}],
\end{equation}
where $\rho_{12}$ is the correlation coefficient of $X_1$ and $X_2$ and $e\sim\mathcal{CN}(0,1)$. Using results from Appendix A,
\begin{equation}
    \mathbb{E}[\nu^*\mathbf{h}_{\mathrm{d}}] = \frac{\sqrt{\beta_\mathrm{d}\pi}\mathbf{R}_{\mathrm{d}}\mathbf{a}_\mathrm{b}}{2\sqrt{\mathbf{a}_\mathrm{b}^\dagger\mathbf{R}_\mathrm{d}\mathbf{a}_\mathrm{b}}}.
\end{equation}
Expanding out all terms in $\mathbf{Z}$ gives
\begin{equation}
    \mathbf{Z}\!=\!\mathbf{R}_\mathrm{b}^{1\!/2}\mathbb{E}[\mathbf{U}_\mathrm{rb}\mathbf{R}_\mathrm{r}^{1/2}\mathrm{diag}(\mathbf{a}_\mathrm{r})\mathbf{B}\mathrm{diag}(\mathbf{a}_\mathrm{r}^*)\mathbf{R}_\mathrm{r}^{1/2}\mathbf{U}_\mathrm{rb}]\mathbf{R}_\mathrm{b}^{1/2}\!, \notag
\end{equation}
where $\mathbf{B} = \mathbb{E}[|\mathbf{h}_\mathrm{ur}||\mathbf{h}_\mathrm{ur}|^T]$. Applying \cite[(7)]{singh_optimal_2021}, the $i,j$-th element of $\mathbf{B}$ is
\begin{equation}
    \mathbf{B}_{i,j} = \beta_\mathrm{ur}\tfrac{\pi}{4}{}_2F_1\left(-\tfrac{1}{2},-\tfrac{1}{2},1,|\mathbf{R}_{\mathrm{ur},i,j}|^2\right). \label{eq:B}
\end{equation}
Finally, for an arbitrary square matrix $\mathbf{X}$, $\mathbb{E}[\mathbf{U^\dagger XU}] = \mathrm{tr}(\mathbf{X})\mathbf{I}$, $\mathbf{Z}$ becomes,
\begin{equation}
    \mathbf{Z} = \mathbf{R}_\mathrm{b}\mathrm{tr}\left(\mathrm{diag}(\mathbf{a}_\mathrm{r})\mathbf{B}\mathrm{diag}(\mathbf{a}_\mathrm{r}^*)\mathbf{R}_\mathrm{r}\right). 
\end{equation}

\section*{Appendix C \\ Derivation of $\mathbb{E}[\mathrm{SNR}(t+\tau)]$}
$\mathbb{E}[\mathrm{SNR}(t+\tau)]$ is required to derive the ageing loss in Section \ref{sec:ageing_loss}, and is given by
\begin{equation}
    \mathbb{E}[\mathrm{SNR}(t+\tau)] = \mathbb{E}[\mathbf{h}^\dagger(t+\tau)\mathbf{h}(t+\tau)],
\end{equation}
and $\mathbf{h}(t+\tau)$ is defined in (\ref{eq:httau}). Expanding out (\ref{eq:httau}) gives,
\begin{multline}
    \mathbf{h}(t+\tau) = \mathbf{h}_\mathrm{d}(t+\tau) + c_1\mathbf{a}_\mathrm{b}\mathbf{a}_\mathrm{r}^\dagger\mathbf{\Phi}(t)\mathbf{h}_\mathrm{ur}(t+\tau) \\ + c_2\mathbf{G}(t+\tau)\mathbf{\Phi}(t)\mathbf{h}_\mathrm{ur}(t+\tau).
\end{multline}
Rewriting $\mathbf{h}_\mathrm{ur}(t+\tau)$ in terms of $\mathbf{h}_\mathrm{ur}(t)$ for the term encompassing the LoS component of $\mathbf{H}_\mathrm{rb}$ in order to simplify the forthcoming expectations and expanding out $\mathbf{\Phi}(t)$, (\ref{eq:httau}) can be written as a sum of four terms, 
\begin{align}
    \mathbf{h}(t\!+\!\tau)\!&=\! \mathbf{h}_\mathrm{d}(t\!+\!\tau) \!+\!c_1\mathbf{a}_\mathrm{b}\nu (t)\Big(\!\rho_\mathrm{ur}(\tau)Y(t)\!+\! \sqrt{1\!-\!|\rho_\mathrm{ur}(\tau)|^2} \notag \\ &\quad\times\!\Big(\!\mathrm{e}^{-j\angle\mathbf{h}_\mathrm{ur}(t)}\!\Big)^T\!\!\mathbf{e}_\mathrm{ur}(t)\Big) +c_2 \mathbf{G}(t\!+\!\tau)\nu (t)\mathrm{diag}(\mathbf{a}_\mathrm{r}) \notag \\ &\quad\times \mathrm{diag}\!\left(\!\mathrm{e}^{-j\angle\mathbf{h}_\mathrm{ur}(t)}\!\right)\!\mathbf{h}_\mathrm{ur}(t\!+\!\tau),\notag \\ 
    & = S_1 + S_2 + S_3 + S_4,
\end{align}
where
\begin{equation}
     S_1 = \mathbf{h}_\mathrm{d}(t+\tau)
\end{equation}
\begin{equation}
     S_2 = c_1\mathbf{a}_\mathrm{b}\nu (t) Y(t)\rho_\mathrm{ur}(\tau)
\end{equation}
\begin{equation}
     S_3 = c_1\sqrt{1-|\rho_\mathrm{ur}(\tau)|^2}\mathbf{a}_\mathrm{b}\nu (t)\left(\mathrm{e}^{-j\angle\mathbf{h}_\mathrm{ur}}(t)\right)^T\!\!\mathbf{e}_\mathrm{ur}(t),
\end{equation}
\begin{equation}
    S_4\!=\!c_2 \mathbf{G}(t\!+\!\tau)\nu (t)\mathrm{diag}(\mathbf{a}_\mathrm{r})\mathrm{diag}\big(\mathrm{e}^{-j\angle\mathbf{h}_\mathrm{ur}(t)}\big)\mathbf{h}_\mathrm{ur}(t\!+\!\tau),
\end{equation}
and $\mathbf{e}_\mathrm{ur}(t)\sim \mathcal{CN}(0,\beta_\mathrm{ur}\mathbf{R}_\mathrm{ur})$.
Note that $\mathbf{h}_\mathrm{d}(t+\tau)$, $\mathbf{h}_\mathrm{ur}(t+\tau)$ and $\mathbf{e}_\mathrm{ur}(t)$ are independent. All three channels are Rayleigh and thus zero mean, so $\mathbb{E}[\mathbf{h}_\mathrm{d}(t+\tau)] = \mathbb{E}[\mathbf{h}_\mathrm{ur}(t+\tau)] = \mathbb{E}[\mathbf{e}_\mathrm{ur}(t)] = 0$. Thus,
\begin{multline}
    \mathbb{E}[\mathrm{SNR}(t+\tau)] = \mathbb{E}[S_1^\dagger S_1] + \mathbb{E}[S_1^\dagger S_2] + \mathbb{E}[S_2^\dagger S_1] \\ + \mathbb{E}[S_2^\dagger S_2] + \mathbb{E}[S_3^\dagger S_3] + \mathbb{E}[S_4^\dagger S_4],
\end{multline}
and each term can be considered independently.
\\

\textit{Term 1}: $\mathbb{E}[S_1^\dagger S_1]$ \\
As this expression only includes $S_1$, time notation can be dropped. From \cite[(20)]{singh_optimal_2021},
\begin{equation}
    \mathbb{E}[S_1^\dagger S_1] = \mathbb{E}[\mathbf{h}_\mathrm{d}^\dagger\mathbf{h}_\mathrm{d}] = M\beta_\mathrm{d}. \label{eq:E_S1S1}
\end{equation}

\textit{Term 2}: $\mathbb{E}[S_1^\dagger S_2]$ \\
Expanding out $S_1$ and $S_2$ and taking the expectation of the independent terms,
\begin{equation}
    \mathbb{E}[S_1^\dagger S_2] = c_1\rho_\mathrm{ur}(\tau)\mathbb{E}[Y(t)]\mathbb{E}[\mathbf{h}_\mathrm{d}^\dagger(t+\tau)\mathbf{a}_\mathrm{b}\nu(t)], \notag
\end{equation}
where $\mathbb{E}[Y(t)]$ is defined in (\ref{eq:EY}) and
\begin{multline}
    \mathbb{E}[\mathbf{h}_\mathrm{d}^\dagger(t+\tau)\mathbf{a}_\mathrm{b}\nu (t)] \\ = \mathbb{E}\left[\left(\rho_\mathrm{d}^*(\tau)\mathbf{h}_\mathrm{d}^\dagger (t) + \sqrt{1\!-\!|\rho_\mathrm{d}(\tau)|^2}\mathbf{e}^\dagger_\mathrm{d}(t)\right)\mathbf{a}_\mathrm{b}\nu (t)\right]. \notag
\end{multline}
As $\mathbf{h}_\mathrm{d}(t)$ and $\mathbf{e}_\mathrm{d}(t)$ are independent, the expression becomes
\begin{align}
    \mathbb{E}[\mathbf{h}_\mathrm{d\!}^\dagger(t{+}\tau)\mathbf{a}_\mathrm{b}\nu (t)]& = \rho^*_\mathrm{d\!}(\tau)\mathbb{E}[|\mathbf{a}_\mathrm{b}^\dagger\mathbf{h}_\mathrm{d}(t)] \notag \\& =\frac{\rho^*_\mathrm{d\!}(\tau)\!\sqrt{\!\beta_\mathrm{d}\pi\mathbf{a}_\mathrm{b}^\dagger\mathbf{R}_\mathrm{d}\mathbf{a}_\mathrm{b}}}{2}. \notag
\end{align}
Therefore,
\begin{equation}
    \mathbb{E}[S_1^\dagger S_2] = \frac{N\pi\eta_\mathrm{rb}\sqrt{\beta_\mathrm{d}\beta_\mathrm{rb}\beta_\mathrm{ur}\mathbf{a}_\mathrm{b}^\dagger\mathbf{R}_\mathrm{d}\mathbf{a}_\mathrm{b}}}{4}\rho^*_\mathrm{d}(\tau)\rho_\mathrm{ur}(\tau). \label{eq:E_S1S2}
\end{equation}

\textit{Term 3}: $\mathbb{E}[S_2^\dagger S_1]$ \\
Following the same method used for Term 2, 
\begin{equation}
    \mathbb{E}[S_2^\dagger S_1] = \frac{N\pi\eta_\mathrm{rb}\sqrt{\beta_\mathrm{d}\beta_\mathrm{rb}\beta_\mathrm{ur}\mathbf{a}_\mathrm{b}^\dagger\mathbf{R}_\mathrm{d}\mathbf{a}_\mathrm{b}}}{4}\rho_\mathrm{ur}^*(\tau)\rho_\mathrm{d}(\tau). \label{eq:E_S2S1}
\end{equation}

\textit{Term 4}: $\mathbb{E}[S_2^\dagger S_2]$ \\
Expanding out $S_2$ gives
\begin{equation}
    \mathbb{E}[S_2^\dagger S_2]\!=\!\mathbb{E}[\rho^*_\mathrm{ur}\!(\tau)Y\!(t)\nu^*\!(t)\mathbf{a}_\mathrm{b}^\dagger c_1^\dagger c_1\mathbf{a}_\mathrm{b}\nu(t)Y\!(t)\rho_\mathrm{ur}(\tau)]. \notag
\end{equation}
Removing constants and given $\mathbb{E}[\mathbf{a}_\mathrm{b}^\dagger\mathbf{a}_\mathrm{b}] = M$ we have
\begin{equation}
    \mathbb{E}[S_2^\dagger S_2]= Mc_1^2|\rho_\mathrm{ur}(\tau)|^2\mathbb{E}[Y^2(t)].
\end{equation}
Therefore, using (\ref{eq:Y2}) for $\mathbb{E}\left[Y^2(t)\right]$ leads to 
\begin{multline}
    \mathbb{E}[S_2^\dagger S_2]= M\beta_\mathrm{rb}\beta_\mathrm{ur}\eta_\mathrm{rb}|\rho_\mathrm{ur}(\tau)|^2\Big(N + \tfrac{\pi}{4} \\ \times\sum_{i=1}^{N}\sum_{j=1}^{N}{}_2F_1\left(-\tfrac{1}{2}, -\tfrac{1}{2},1;|\mathbf{R}_{\mathrm{ur},i,j}|^2\right)\!\Big). \label{eq:E_S2S2}
\end{multline}

\textit{Term 5}: $\mathbb{E}[S_3^\dagger S_3]$ \\
Expanding out $S_3$ and simplifying,
\begin{multline}
    \mathbb{E}[S_3^\dagger S_3] = M\beta_\mathrm{rb}\eta_\mathrm{rb}^2(1-|\rho_\mathrm{ur}(\tau)|^2)\\ \times\mathbb{E}\Big[\!\Big(\!\mathrm{e}^{-j\angle\mathbf{h}_\mathrm{ur}(t)}\!\Big)^{\!T}\!\mathbf{e}_\mathrm{ur}(t)\mathbf{e}_\mathrm{ur}^\dagger(t)\mathrm{e}^{j\angle\mathbf{h}_\mathrm{ur}(t)}\Big]. \label{eq:ES3s3_expec}
\end{multline}
As $\mathbf{e}_\mathrm{ur}$(t) is a correlated Rayleigh channel with correlation matrix $\mathbf{R}_\mathrm{ur}$,
\begin{equation}
    \mathbb{E}[\mathbf{e}_\mathrm{ur}(t)\mathbf{e}_\mathrm{ur}(t)^\dagger] = \beta_\mathrm{ur}\mathbf{R}_\mathrm{ur}. \label{eq:eureurdag}
\end{equation}
Therefore, the expectation in (\ref{eq:ES3s3_expec}) can be rewritten as
\begin{multline}
    \mathbb{E}[S_3^\dagger S_3] = M\beta_\mathrm{rb}\beta_\mathrm{ur}\eta_\mathrm{rb}^2(1-|\rho_\mathrm{ur}(\tau)|^2)\sum_{i=1}^N\sum^N_{j=1}\mathbf{R}_{\mathrm{ur},i,j} \\ \times\mathbb{E}\big[\mathrm{e}^{j(\angle\mathbf{h}_{\mathrm{ur},j}(t) - \angle\mathbf{h}_{\mathrm{ur},i}(t))}\big].
\end{multline}
Using the result in (30) of \cite{inwood_phase_2023},
\begin{multline}
    \mathbb{E}[S_3^\dagger S_3] = M\beta_\mathrm{rb}\beta_\mathrm{ur}\eta_\mathrm{rb}^2\tfrac{\pi}{4}(1-|\rho_\mathrm{ur}(\tau)|^2)\sum_{i=1}^N\sum^N_{j=1}(\mathbf{R}_{\mathrm{ur},i,j})^2 \\ \times {}_2F_1\left(\tfrac{1}{2},\tfrac{1}{2},2,|\mathbf{R}_{\mathrm{ur},i,j}|^2\right). \label{eq:E_S3S3}
\end{multline}

\textit{Term 6}: $\mathbb{E}[S_4^\dagger S_4]$ \\
Expanding out $S_4$,
\begin{multline*}
    \mathbb{E}[S_4^\dagger S_4]\!=\!c_2^2\mathbb{E}\big[\mathbf{h}^\dagger_\mathrm{ur}(t\!+\!\tau)\mathrm{diag}\big(\mathrm{e}^{j\angle\mathbf{h}_\mathrm{ur}(t)}\big)\mathrm{diag}(\mathbf{a}_\mathrm{r}^*)\nu^*\!(t) \\ \times\!\mathbf{G}^\dagger\!(t\!+\!\tau)
    \mathbf{G}(t\!+\!\tau)\nu(t)\mathrm{diag}(\mathbf{a}_\mathrm{r})\mathrm{diag}\big(\mathrm{e}^{-j\angle\mathbf{h}_\mathrm{ur}(t)}\big)\mathbf{h}_\mathrm{ur}(t\!+\!\tau)\!\big]\!.
\end{multline*}
Again, for an arbitrary square matrix $\mathbf{X}$, as $\mathbb{E}[\mathbf{U^\dagger X U}] = \mathrm{tr}\mathbf{(X)I}$, $\mathbb{E}[\mathbf{G}^\dagger(t+\tau)\mathbf{G}(t+\tau)]$ $= \mathbb{E}[\mathbf{R}_\mathrm{r}^{1/2}\mathbf{U}^\dagger_\mathrm{rb}\mathbf{R}_\mathrm{b}^{1/2}\mathbf{R}_\mathrm{b}^{1/2}\mathbf{U}_\mathrm{rb}\mathbf{R}_\mathrm{r}^{1/2}]$ becomes $M\mathbf{R}_\mathrm{r}$. Therefore, using this result and rewriting $\mathbf{h}_\mathrm{ur}(t+\tau)$ in terms of $t$ gives
\begin{align}
    \mathbb{E}[&S_4^\dagger S_4] = Mc_2^2\mathbb{E}\big[\!\big(\rho_\mathrm{ur}^*(\tau)\mathbf{h}_\mathrm{ur}^\dagger(t)+\sqrt{1-|\rho_\mathrm{ur}(\tau)|}\mathbf{e}^\dagger_\mathrm{ur}(t)\!\big) \notag \\ &\times\! \mathrm{diag}\big(\!\mathrm{e}^{j\angle\mathbf{h}_\mathrm{ur}(t)}\big)\mathrm{diag}(\mathbf{a}_\mathrm{r}^*)\mathbf{R}_\mathrm{r}\mathrm{diag}(\mathbf{a}_\mathrm{r})\mathrm{diag}\big(\!\mathrm{e}^{-j\angle\mathbf{h}_\mathrm{ur}(t)}\!\big)\notag \\ &\times\big(\rho_\mathrm{ur}(t) \mathbf{h}_\mathrm{ur}(t)+\sqrt{1-|\rho_\mathrm{ur}(\tau)|}\mathbf{e}_\mathrm{ur}(t)\!\big)\big].
\end{align}
Expanding out the brackets, and knowing that $\mathbf{h}_\mathrm{ur}(t)$ and $\mathbf{e}_\mathrm{ur}(t)$ are independent, expectations including both terms are equal to 0. Using this, the fact that $\mathbf{e}_\mathrm{ur}(t) \sim \mathcal{CN}(0,\beta_\mathrm{ur}\mathbf{R}_\mathrm{ur})$ and the identity $\mathbb{E}[\mathbf{U^\dagger X U}] = \mathrm{tr}\mathbf{(X)I}$, 
\begin{align}
    &\mathbb{E}[S_4^\dagger S_4] = Mc_2^2\big(|\rho_\mathrm{ur}(\tau)|^2\mathrm{tr}\big(\mathrm{diag}(\mathbf{a}_\mathrm{r}^*)\mathbf{R}_\mathrm{r}\mathrm{diag}(\mathbf{a}_\mathrm{r})\notag \\ &\times\!\mathbb{E}\!\left[|\mathbf{h}_\mathrm{ur}(t)||\mathbf{h}_\mathrm{ur}(t)|^T\right]\!\big)\!+\!\left(1\!-\!|\rho_\mathrm{ur}(\tau)|^2\right)\mathrm{tr}(\mathbf{R}_\mathrm{ur}\mathbf{D})\!\big),
\end{align}
where
\begin{equation}
    \!\mathbf{D}\!=\!\mathbb{E}\big[\mathrm{diag}\big(\!\mathrm{e}^{j\angle\mathbf{h}_\mathrm{ur}(t)}\!\big)\mathrm{diag}(\mathbf{a}_\mathrm{r}^*)\mathbf{R}_\mathrm{r}\mathrm{diag}(\mathbf{a}_\mathrm{r})\mathrm{diag}\big(\!\mathrm{e}^{-j\angle\mathbf{h}_\mathrm{ur}(t)}\!\big)\!\big]\!. \notag
\end{equation}
Each element in D can be written as 
\begin{equation}
    \mathbf{D}_{i,j} = \mathbb{E}\big[\mathrm{e}^{j(\angle\mathbf{h}_{\mathrm{ur},i}(t) - \angle\mathbf{h}_{\mathrm{ur},j}(t))}\big]\mathbf{R}_{\mathrm{r},i,j}\,\mathbf{a}_{\mathrm{r},i}^*\,\mathbf{a}_{\mathrm{r},j},
\end{equation}
which, using the result in (30) of \cite{inwood_phase_2023}, is
\begin{equation}
    \!\mathbf{D}_{i,j}\!=\!\tfrac{\pi}{4}\mathbf{R}_{\mathrm{r},i,j}\mathbf{R}_{\mathrm{ur},i,j}\mathbf{a}_{\mathrm{r},i}^*\,\mathbf{a}_{\mathrm{r},j}\,{}_2F_1\!\left(\tfrac{1}{2},\tfrac{1}{2},2;|\mathbf{R}_{\mathrm{ur},i,j}|^2\right)\!.\!
\end{equation}
Therefore, using the result in (\ref{eq:B}) for $\mathbb{E}\!\left[|\mathbf{h}_\mathrm{ur}(t)||\mathbf{h}_\mathrm{ur}(t)|^T\right]$,
\begin{align}
    &\mathbb{E}[S_4^\dagger S_4]\!=\!M\beta_\mathrm{rb}\beta_\mathrm{ur}\zeta_\mathrm{rb}^2\big(\!|\rho_\mathrm{ur}(\tau)|^2\mathrm{tr}\big(\tfrac{\pi}{4}\mathrm{diag}(\mathbf{a}_\mathrm{r}^*)\mathbf{R}_\mathrm{r}\mathrm{diag}(\mathbf{a}_\mathrm{r})\notag \\ &\times\!{}_2F_1\!\left(\!-\tfrac{1}{2},-\tfrac{1}{2},\!1,\!|\mathbf{R}_\mathrm{ur}|^2\right)\!\!\big)\!\!+\!\!\left(1\!-\!|\rho_\mathrm{ur}(\tau)|^2\right)\!\mathrm{tr}\big(\mathbf{R}_\mathrm{ur}\mathbf{D}\big)\!\big)\!.\!\! \label{eq:E_S4S4}
\end{align}
Combining (\ref{eq:E_S1S1}), (\ref{eq:E_S1S2}), (\ref{eq:E_S2S1}), (\ref{eq:E_S2S2}), (\ref{eq:E_S3S3}) and (\ref{eq:E_S4S4}) gives (\ref{eq:ESNRtplustau}).

\vspace{-10pt}
\bibliographystyle{IEEEtran}
\bibliography{IEEEabrv, journal.bib}
\vspace{-35pt}
\begin{IEEEbiography}[{\includegraphics[width=1in,height=1.25in,clip,keepaspectratio]{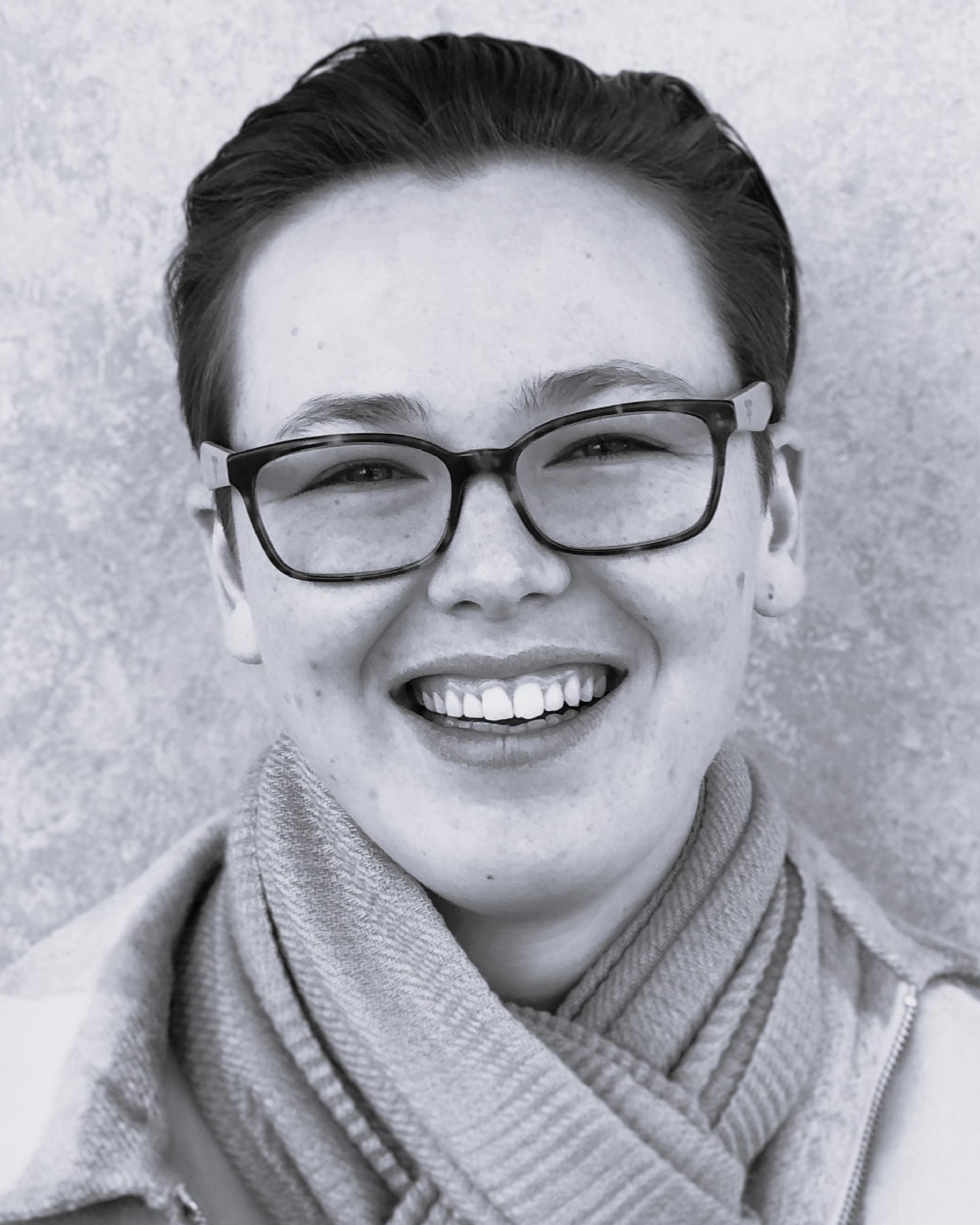}}]%
{Amy S. Inwood} (S'18-M'25) received the B.E (Hons.) and Ph.D. degrees in electrical and electronic engineering from Te Whare Wānanga o Waitaha $|$ University of Canterbury (UC), NZ, in 2021 and 2024, respectively. She is now a Research Fellow at the Centre for Wireless Innovation, Queen's University Belfast, Belfast, U.K. Her research interests include statistical analysis, 5G-6G wireless communications, MIMO, reconfigurable intelligent surfaces and fluid antenna systems.
\end{IEEEbiography}
\vspace{-30pt}
\begin{IEEEbiography}[{\includegraphics[width=1in,height=1.25in,clip,keepaspectratio]{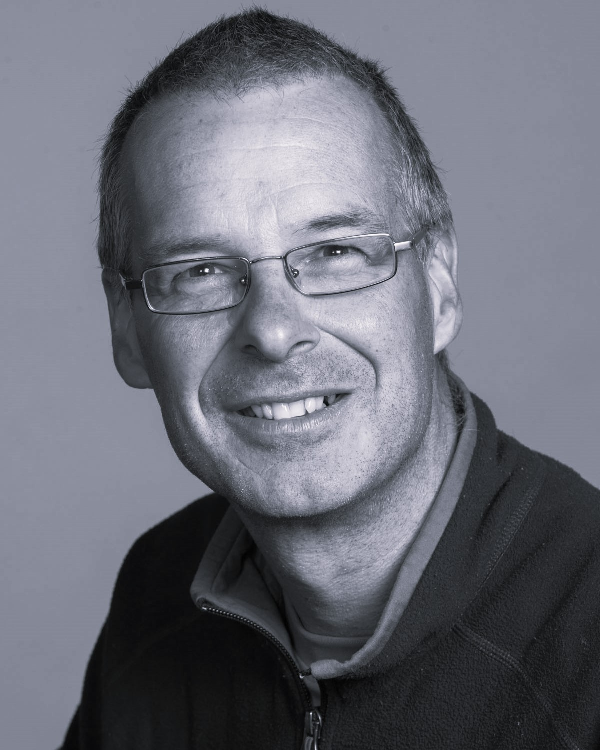}}]%
{Peter J. Smith} (M'93-SM'01-F'15) received the B.Sc degree in Mathematics and the Ph.D degree in Statistics from the University of London, London, U.K., in 1983 and 1988, respectively. In 2015 he joined Victoria University of Wellington as Professor of Statistics. He is an Adjunct Professor in Electrical and Computer Engineering at the University of Canterbury, New Zealand and an Honorary Professor in the School of Electronics, Electrical Engineering and Computer Science, Queens University Belfast. His research interests include the statistical aspects of design, antenna arrays, MIMO, cognitive radio, massive MIMO, mmWave systems, reconfigurable intelligent surfaces and the fusion of radar sensing and communications.
\end{IEEEbiography}
\vspace{-30pt}
\begin{IEEEbiography}[{\includegraphics[width=1in,height=1.25in,clip,keepaspectratio]{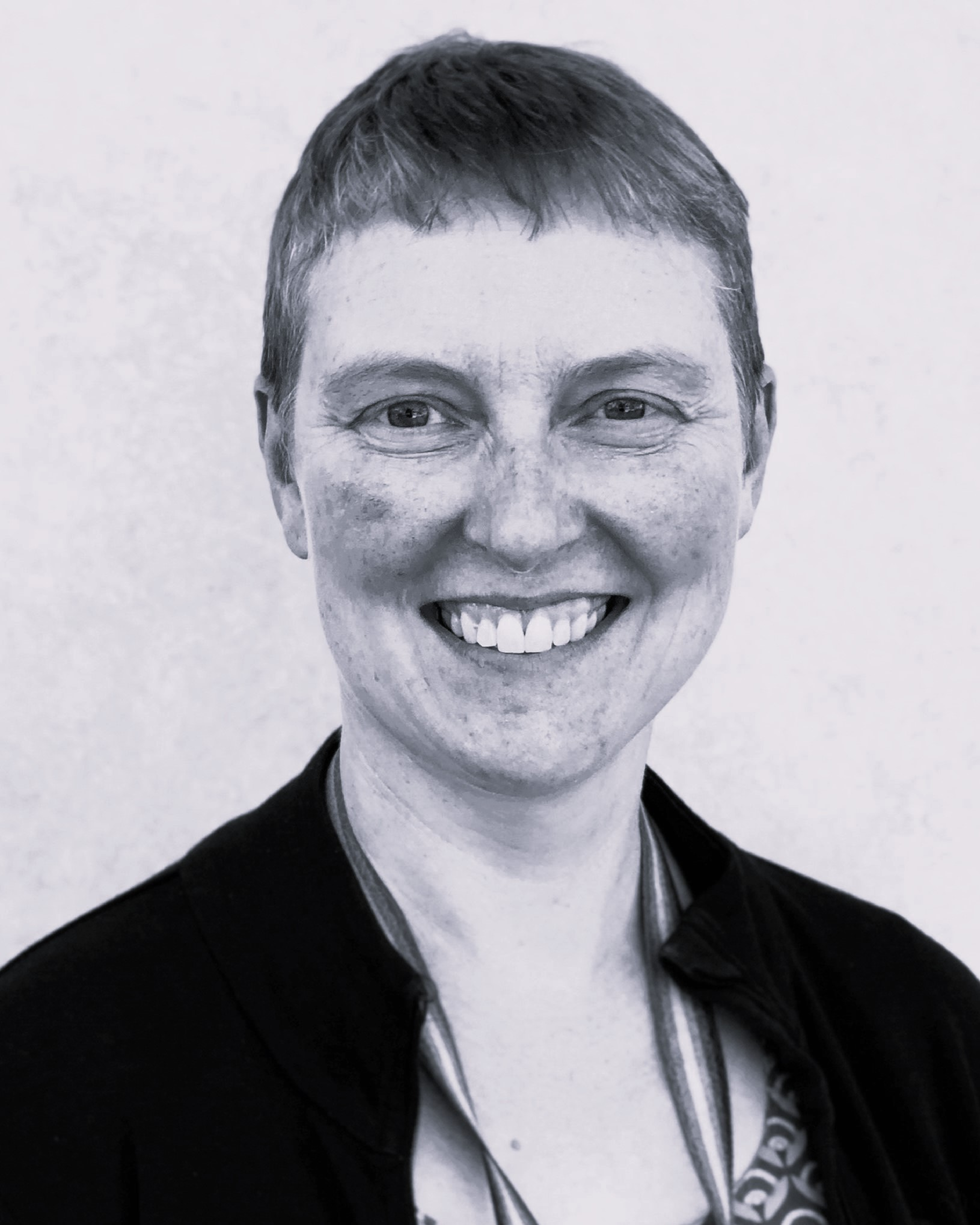}}]%
{Philippa A. Martin} (S'95-M'01-SM'06) received the B.E. (Hons.) and Ph.D. degrees in electrical and electronic engineering from Te Whare Wānanga o Waitaha $|$ University of Canterbury, NZ, in 1997 and 2001, respectively. She is now a Professor there. She is a Fellow of Engineering New Zealand. Her research interests include error correction coding, detection, multi-antenna systems, channel modelling, and 5G-6G wireless communications. She served as an Editor of the IEEE Transactions on Wireless Communications 2005-08, 2014-16 and member of the IEEE Communication Society Board of Governors 2019-21. She is currently on their financial standing committee.
\end{IEEEbiography}
\vspace{-30pt}
\begin{IEEEbiography}[{\includegraphics[width=1in,height=1.25in,clip,keepaspectratio]{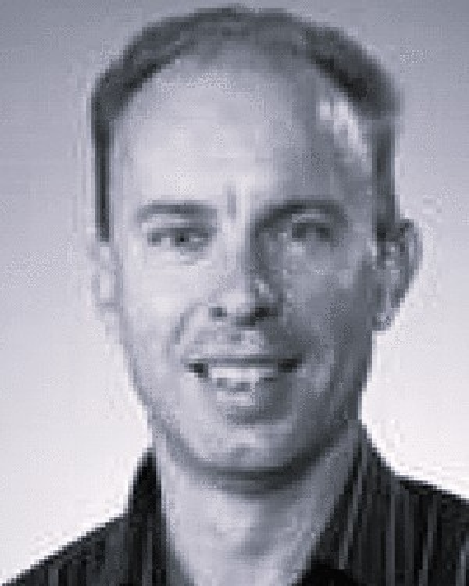}}]{Graeme K. Woodward} (S'94-M'99-SM'05) received the B.Sc., B.E., and Ph.D. degrees from The University of Sydney, NSW, Australia, in 1990, 1992, and 1999, respectively. He has been the Research Leader with the Wireless Research Centre, Te Whare Wānanga o Waitaha $|$ University of Canterbury, NZ, since 2011, and previously the Research Manager of the Telecommunications Research Laboratory, Toshiba Research Europe, contributing to numerous large U.K. and EU projects. His extensive industrial research experience includes pioneering VLSI designs for multi-antenna 3G Packet Access (HSDPA) with Bell Labs Sydney. While holding positions at Agere Systems and LSI Logic, his focus moved to terminal-side algorithms for 3G and 4G (LTE), with an emphasis on low power design. \end{IEEEbiography}

\end{document}